\documentstyle[12pt]{article}

\def\bs{\bigskip}
\def\ms{\medskip}
\def\cl{\centerline}
\def\ref{\par\smallskip\hangindent=1.0cm\hangafter=1}

\begin{document}

\baselineskip=14pt plus 1pt minus 1pt 

\cl{\bf STAGGERING EFFECTS IN NUCLEAR AND MOLECULAR SPECTRA}

\bs
\cl{DENNIS BONATSOS, N. KAROUSSOS}

\cl{Institute of Nuclear Physics, N.C.S.R. ``Demokritos'',}

\cl{GR-15310 Aghia Paraskevi, Attiki, Greece}

\cl{C. DASKALOYANNIS}

\cl{Department of Physics, Aristotle University of
Thessaloniki,}

\cl{GR-54006 Thessaloniki, Greece}

\cl{S. B. DRENSKA, N. MINKOV, P. P. RAYCHEV, R. P. ROUSSEV}

\cl{Institute for Nuclear Research and Nuclear Energy, Bulgarian
Academy of Sciences,}

\cl{72 Tzarigrad Road, BG-1784 Sofia, Bulgaria}

\cl{J. MARUANI}

\cl{Laboratoire de Chimie Physique, CNRS and UPMC,}

\cl{11, rue Pierre et Marie Curie, F-75005 Paris, France}

\bs\bs
\cl{\bf Abstract}
\ms

It is  shown that the recently observed
$\Delta J=2$ staggering effect (i.e. the relative displacement of the
levels with angular momenta $J$, $J+4$, $J+8$, \dots, relatively to the
levels with angular momenta $J+2$, $J+6$, $J+10$, \dots) seen in superdeformed
nuclear bands is also occurring in certain electronically excited rotational
bands of diatomic molecules (YD, CrD, CrH, CoH), in which it is attributed to
interband interactions (bandcrossings). In addition, the $\Delta J=1$
staggering effect (i.e. the relative displacement of the levels
with even angular momentum $J$ with respect to the levels of the same band
with odd $J$) is studied in molecular bands free from $\Delta J=2$ staggering
(i.e. free from interband interactions/bandcrossings). Bands of YD offer
evidence for the absence of any $\Delta J=1$ staggering effect due to the
disparity of nuclear masses, while bands of sextet electronic states of CrD
demonstrate that $\Delta J=1$ staggering is a sensitive probe of deviations
from rotational behaviour, due in this particular case to the
spin--rotation and spin--spin interactions.

\bs\bs

\section{Introduction}

Several {\sl staggering} effects are known in nuclear spectroscopy \cite{BM}:

1) In rotational $\gamma$ bands of even nuclei the energy levels with
odd angular momentum $I$ ($I$=3, 5, 7, 9, \dots) are slightly displaced
relatively to the levels with even $I$ ($I$=2, 4, 6, 8, \dots), i.e.
the odd levels do not lie at the energies predicted by an $E(I)=A I(I+1)$ fit
to the even levels, but all of them lie systematically above or all of
them lie systematically below the predicted energies \cite{PLB200}.

2) In octupole bands of even nuclei the levels with odd $I$ and negative
parity ($I^{\pi}$=1$^-$, 3$^-$, 5$^-$, 7$^-$, \dots) are displaced relatively
to the levels with even $I$ and positive parity ($I^{\pi}$=0$^+$, 2$^+$,
4$^+$, 6$^+$, \dots) \cite{Phill,Schuler,Ahmad,Butler}.

3) In odd nuclei, rotational bands (with $K=1/2$) separate into
signature partners, i.e. the levels with $I$=3/2, 7/2, 11/2, 15/2, \dots
are displaced relatively to the levels with $I$=1/2, 5/2, 9/2, 13/2, \dots
\cite{WuZhou}.

In all of the above mentioned cases each level with angular momentum $I$
is displaced relatively to its neighbours with angular momentum $I\pm 1$.
The effect is then called {\sl $\Delta I=1$ staggering}. In all cases
the effect has been seen in several nuclei and its magnitude is clearly
larger than the experimental errors. In cases 1) and 3) the relative
displacement of the neighbours increases in general as a function of the
angular momentum $I$ \cite{PLB200,WuZhou},
while in case 2) (octupole bands), the relevant models
\cite{EI1126,EI61,GRR8,GRR9,GRR12}
predict constant displacement of the odd levels with respect
to the even levels as a function of $I$, i.e. all the odd levels
are raised (or lowered) by the same amount of energy.

A new kind of staggering ({\sl $\Delta I=2$ staggering}) has been recently
observed \cite{Fli,Ced} in superdeformed nuclear bands
\cite{Twin,Nolan,Janssens}.
In the case in which $\Delta I=2$ staggering is present,
the levels with $I$=2, 6, 10, 14, \dots,
for example,  are displaced
relatively to the levels with $I$=0, 4, 8, 12, \dots, i.e. the level with
angular momentum $I$ is displaced relatively to its neighbours with
angular momentum $I\pm 2$.

Although $\Delta I=1$ staggering of the types mentioned above has been
observed in several nuclei and certainly is an effect larger than the
relevant experimental uncertainties, $\Delta I=2$ staggering has been seen
in only a few cases \cite{Fli,Ced,Semple,Kruecken}
and, in addition, the effect is not clearly larger
than the relevant experimental errors.

There have been by now several theoretical works related to the
possible physical origin of the $\Delta I=2$ staggering effect
\cite{SZG,MQ,Mag,Kota,Liu,Pav,Wu},
some of them \cite{HM,Macc,PavFli,Doenau,Luo,Magi}
using symmetry arguments which could be of applicability
to other physical systems as well.

On the other hand, rotational spectra of diatomic molecules
\cite{Herz} are known to
show great similarities to nuclear rotational spectra, having in addition
the advantage that observed  rotational bands in several diatomic molecules
\cite{YD,CrD,CrH,CoH}
are much longer than the usual rotational nuclear bands. We have been
therefore motivated to make a search for $\Delta J=1$ and
$\Delta J =2$ staggering in rotational bands of diatomic molecules,
where by $J$ we denote the total angular momentum of the molecule, while $I$
has been used above for denoting the angular momentum of the nucleus.
The questions to which we have hoped
to provide answers are:

1) Is there $\Delta J=1$ and/or $\Delta J =2$ staggering in rotational bands
of diatomic molecules?

2) If there are staggering effects, what are their possible physical origins?

In Sections 2 and 3 the $\Delta J=2$ staggering and $\Delta J=1$
staggering will be considered respectively, while in Section 4 the final
conclusions and plans for further work will be presented.

\section{$\Delta J=2$ staggering} 

In this section the $\Delta J=2$ staggering will be considered. In subsection
2.1 the $\Delta I=2$ staggering in superdeformed nuclear bands will be
briefly reviewed. Evidence from existing experimental data for $\Delta J=2$
staggering in rotational bands of diatomic molecules will be presented in
subsection 2.2 and discussed in subsection 2.3, while subsection 2.4 will
contain the relevant conclusions. We mention once more that by $J$ we denote
the total angular momentum of the molecule, while by $I$ the angular
momentum of the nucleus is denoted.

\subsection{$\Delta I=2$ staggering in superdeformed nuclear bands} 

In nuclear physics the experimentally determined quantities are the
$\gamma$-ray transition energies between levels differing by two units
of angular momentum ($\Delta I=2$). For these the symbol
\begin{equation}
E_{2,\gamma}(I) = E(I+2)-E(I)
\end{equation} 
is used, where $E(I)$ denotes the energy of the level with angular momentum
$I$.
The deviation of the $\gamma$-ray transition energies from the
rigid rotator behavior can be measured by the quantity \cite{Ced}
$$ \Delta E_{2,\gamma}(I) = {1\over 16} (6E_{2,\gamma}(I) -4E_{2,\gamma} (I-2)
-4E_{2,\gamma}(I+2)  $$ 
\begin{equation}
+E_{2,\gamma}(I-4) +E_{2,\gamma}(I+4)).
\end{equation} 

\noindent Using the rigid rotator expression
\begin{equation}
E(I)=A I(I+1),
\end{equation}
one can easily see that
in this case $\Delta E_{2,\gamma} (I) $ vanishes.
In addition the perturbed rigid rotator expression
\begin{equation}
E(I)= A I(I+1) + B (I(I+1))^2,
\end{equation}
 gives vanishing $\Delta E_{2,\gamma} (I)$.
These properties are due to the fact that Eq. (2) is a (normalized)
discrete approximation of the fourth derivative of the function
$E_{2,\gamma}(I)$, i.e. essentially the fifth derivative of the
function $E(I)$.

In superdeformed nuclear bands the angular momentum of the observed states
is in most cases unknown. To avoid this difficulty, the quantity
$\Delta E_{2,\gamma}$ is usually plotted not versus the angular momentum $I$,
but versus the angular frequency
\begin{equation}
\omega = {dE(I)\over dI},
\end{equation}
which for discrete states takes the approximate form
\begin{equation}
\omega = {E(I+2)-E(I)\over \sqrt{(I+2)(I+3)}-\sqrt{I(I+1)} }.
\end{equation}
For large $I$ one can take the Taylor expansions of the square roots in
the denominator, thus obtaining
\begin{equation}
\omega = {E(I+2)-E(I) \over 2} = {E_{2,\gamma}(I) \over 2}.
\end{equation}

Examples of superdeformed nuclear bands exhibiting staggering are shown in
Figs 1--2 \cite{Fli,Ced}. We say that $\Delta I=2$ staggering is observed if
the quantity $\Delta E_2(I)$ exhibits alternating signs with increasing
$\omega$ (i.e. with increasing $I$, according to Eq. (7)). The following
observations can be made:

1) The magnitude of $\Delta E_2(I)$ is of the order of 10$^{-4}$--10$^{-5}$
times the size of the gamma transition energies.

2) The best example of $\Delta I=2$ staggering is given by the first
superdeformed band of $^{149}$Gd, shown in Fig. 1a. In this case the effect
is almost larger than the experimental error.

3) In most cases the $\Delta I=2$ staggering is smaller than the experimental
error (see Figs 1b, 2a, 2b), with the exception of a few points in Fig. 1b.

\subsection{$\Delta J=2$ staggering in rotational bands of diatomic molecules}

In the case of molecules \cite{Bar} the experimentally determined
quantities regard the R branch ($(v_{lower},J)\rightarrow (v_{upper},J+1)$)
and the P branch ($(v_{lower},J)\rightarrow (v_{upper},J-1)$), where
$v_{lower}$ is the vibrational quantum number of the initial state,
while $v_{upper}$ is the vibrational quantum number of the final state.
They are related to transition energies through the equations \cite{Bar}
\begin{equation}
E^R(J)-E^P(J)= E_{v_{upper}} (J+1) -E_{v_{upper}} (J-1) =
DE_{2, v_{upper}} (J),
\end{equation} 
\begin{equation}
E^R(J-1)-E^P(J+1) = E_{v_{lower}}(J+1)-E_{v_{lower}}(J-1)=
DE_{2, v_{lower}}(J),
\end{equation} 
where in general
\begin{equation}
 DE_{2,v} (J) = E_v(J+1)-E_v(J-1).
\end{equation} 
$\Delta J=2$ staggering  can then
be estimated by using Eq. (2), with $E_{2,\gamma}(I)$ replaced by
$DE_{2,v}(J)$:
$$  \Delta E_{2,v} (J)= {1\over 16} (6 DE_{2,v}(J)-4 DE_{2,v}(J-2)
-4 DE_{2,v}(J+2) $$
\begin{equation}
+DE_{2,v}(J-4) +DE_{2,v}(J+4)).
\end{equation} 

Results for several rotational bands in different electronic and vibrational
states of various diatomic molecules are shown in Figs 3--9.
We say that $\Delta J=2$ staggering is observed if the quantity
$\Delta E_2(J)$ exhibits alternating signs with increasing $J$ ($J$ is
increased by 2 units each time). The magnitude of $\Delta E_2(J)$ is
usually of the order of 10$^{-3}$--10$^{-5}$ times the size of the
interlevel separation energy. In Figs 7 and 8, which correspond to sextet
electronic states, the rotational angular momentum $N$ is used instead
of the total angular momentum $J$, the two quantities been connected
by the relation ${\bf J}= {\bf N} +{\bf S}$, where $S$ is the spin.
Several observations can be made:

1) In all cases shown, the ``upper'' bands (which happen
to be electronically
excited) exhibit (Figs 3, 4, 7-9)
$\Delta J=2$ staggering (or $\Delta N=2$ staggering)
which is 2 to 3 orders of magnitude
larger than the experimental error, while
the corresponding ``lower'' bands (which, in the cases studied, correspond to
the electronic ground state of each molecule), show (Figs 5, 6)
some effect smaller than the experimental error.

2) There is no uniform dependence of the $\Delta J=2$ staggering on the
angular momentum $J$. In some cases of long bands, though, it appears that
the pattern is a sequence of points exhibiting small staggering,
interrupted by groups of 6 points each time showing large staggering.
The best examples can be seen in Figs  3a, 3b, 7a, 7b. In Fig. 3a
(odd levels of the $v=1$ C$^1 \Sigma ^+$ band of YD))
the first group of points showing appreciable $\Delta J =2$
staggering appears at $J=13$--23, while the second group appears at
$J=27$--37. In Fig. 3b (even levels of the $v=1$ C$^1 \Sigma^+$ band of YD)
the first group appears at $J=12$--22,
while the second group at $J=26$--36. In Fig. 7a (odd levels of the $v=0$
A$^6 \Sigma ^+$ band of CrD)
the first group
appears at $N=15$--25, while the second at $N=27$--37. Similarly
in Fig. 7b (even levels of the $v=0$ A$^6 \Sigma ^+$ band of CrD)
the first group appears at $N=14$--24, while the second
group at $N=26$--36.

3) In all cases shown, the results obtained for the odd levels of a band
are in good agreement with the results obtained for the even levels of the
same band. For example, the regions showing appreciable staggering
are approximately the same in both cases (compare Fig. 3a with Fig. 3b
and Fig. 7a with Fig. 7b, already discussed in 2)~).  In addition, the
positions of the local staggering maxima in each pair of figures are
closely related. In Fig. 3a, for example, maximum staggering appears at
$J=19$ and $J=31$, while in Fig. 3b the maxima appear at $J=18$ and
$J=32$.

4) In several cases the $\Delta J=2$ staggering of a band can be calculated
from two different sets of data. For example, Figs 3a, 3b show the
$\Delta J=2$ staggering of the $v=1$ C$^1 \Sigma^+$ band of YD calculated
from the data on the 1--1 C$^1 \Sigma ^+$--X$^1\Sigma^+$ transitions,
while Figs 3c, 3d show the staggering of the same band calculated from the
data on the 1--2 C$^1 \Sigma ^+$--X$^1 \Sigma^+$ transition.
We remark that the results
concerning points showing staggering larger than the experimental error
come out completely consistently from the two calculations (region
with $J=13$--23 in Figs 3a, 3c; region with $J=12$--22 in Figs 3b, 3d),
while the results concerning points exhibiting staggering of the order
of the experimental error come out randomly (in Fig. 3a, for example,
$J=11$ corresponds to a local minimum, while in Fig. 3c it corresponds to
a local maximum). Similar results are seen in the pairs of figures
(3b, 3d), (4a, 4c), (4b, 4d), (6a, 6c), (6b, 6d), (9a, 9c), (9b, 9d).
The best example of disagreement
between staggering pictures of the
same band calculated from two different sets of data is offered
by Figs 6b, 6d, which concern the $v=2$ X$^1 \Sigma ^+$ band of YD,
which shows staggering of the order of the experimental error.

5) When considering levels of the same band, in some cases the odd levels
exhibit larger staggering than the even levels, while in other cases the
opposite is true. In the $v=1$ C$^1\Sigma^+$ band of YD, for example,
the odd levels (shown in Fig. 3a, corroborated by Fig. 3c) show staggering
larger than that of the even levels (shown in Fig. 3b, corroborated by
Fig. 3d), while in the $v=2$ C$^1\Sigma^+$ band of YD the odd levels
(shown in Fig. 4a, corroborated by Fig. 4c) exhibit staggering smaller
than that of the even levels (shown in Fig. 4b, corroborated by Fig. 4d).

\subsection{Discussion} 

The observations made above can be explained by the assumption that
the staggering observed is due to the presence of one or more bandcrossings
\cite{Pavli,MRM}.
The following points support this assumption:

1) It is known \cite{VDS}  that bandcrossing occurs in cases in which
the interband interaction is weak. In such cases only the one or two levels
closest to the crossing point are affected \cite{Bonbb}.
However, if one level is influenced
by the crossing, in the corresponding staggering figure six points get
influenced. For example, if E(16) is influenced by the crossing,
the quantities $DE_2(15)$ and $DE_2(17)$ are influenced (see Eq. (10)~),
so that in the corresponding figure the points $\Delta E_2(J)$ with
$J=11$, 13, 15, 17, 19, 21 are influenced,  as seen from Eq. (11).
This fact explains why points showing
appreciable staggering appear in groups of 6 at a time.

2) It is clear that if bandcrossing occurs, large staggering should appear
in approximately
the same angular momentum regions of both even levels and odd levels.
As we have already seen, this is indeed the case.

3) It is clear that when two bands cross each other, maximum staggering
will appear at the angular momentum for which the energies of the relevant
levels of each band are approximately equal \cite{Bonbb}.
If this angular momentum
value happens to be odd, then $\Delta E_2(J)$ for even values of $J$
 in this region (the group
of 6 points centered at this $J$) will show larger staggering than the
$\Delta E_2(J)$ for odd values of $J$ in the corresponding region,
and vice versa. For example,
if the closest approach of two bands occurs for $J=31$, then $\Delta E_2(J)$
for even values of $J$ in the $J=26$--36 region will show larger staggering
than $\Delta E_2(J)$ for odd values of $J$ in the same region. This is in
agreement with the empirical
observation that in some cases the odd levels show larger staggering than
the even levels, while in other cases the opposite holds.

4) The presence of staggering in the ``upper'' (electronically excited)
bands and the lack of staggering in the ``lower'' (electronic ground
state) bands can be attributed to the fact that the electronically
excited bands have several neighbours with
which they can interact, while the bands built on the electronic
ground state are relatively isolated, and therefore no bandcrossings
occur in this case. In the case of the CrD molecule, in particular,
it is known \cite{CrD} that there are many strong Cr atomic lines
present, which frequently overlap the relatively weaker
(electronically excited) molecular lines. In addition, Ne atomic lines
are present \cite{CrD}. Similarly, in the case of the
YD molecule the observed spectra are influenced by Y and Ne atomic lines
\cite{YD}, while in the case of the CrH molecule there are Ne and Cr
atomic lines influencing the molecular spectra \cite{CrH}.

5) The fact that consistency between results for the same band calculated
from two different sets of data is observed only in the cases in which
the staggering is much larger than the experimental error, corroborates
the bandcrossing explanation. The fact that the results obtained in areas
in which the staggering is of the order of the experimantal error, or
even smaller, appear to be random, points towards the absence of any
real effect in these regions.

It should be noticed that bandcrossing has been proposed \cite{RJR,HS,HL}
as a possible explanation for the appearance of $\Delta I=2$ staggering
effects in normally deformed nuclear bands \cite{Wu,RJR,HL} and
superdeformed nuclear bands \cite{HS}.

The presence of two subsequent bandcrossings can also provide an explanation
for the effect of mid-band disappearance of $\Delta I=2$ staggering
observed in superdeformed bands of some Ce isotopes \cite{Semple}.
The effect seen in the Ce isotopes is very similar to the mid-band
disappearance of staggering seen, for example, in Fig. 3a.

\subsection {Conclusion} 

In conclusion, we have found several examples of $\Delta J=2$ staggering
in electronically excited bands of diatomic molecules. The details of
the observed effect are in agreement with the assumption that it is due
to one or more bandcrossings. In these cases the magnitude of the effect
is clearly larger than the experimental error. In cases in which
an effect of the order of the experimental error appears, we have shown
that this is an artifact of the method used, since different sets of data
from the same experiment and for the same molecule lead to different
staggering results for the same rotational band. The present work
emphasizes the need to ensure in all cases (including staggering candidates
in nuclear physics) that the effect is larger than the experimental
error and, in order to make assumptions about any new symmetry,
that it is not due to a series of bandcrossings.

\section{$\Delta J=1$ staggering} 

In this section the
$\Delta J=1$ staggering effect (i.e. the relative displacement of the levels
with even angular momentum $J$ with respect to the levels of the same band
with odd $J$) will be considered in molecular bands free from $\Delta J=2$
staggering (i.e. free from interband interactions/bandcrossings),
in order to make sure that $\Delta J=1$
staggering is not an effect due to the same cause as $\Delta J=2$
staggering.

The formalism of the $\Delta J=1$ staggering
will be described in subsection 3.1 and  applied to
experimental molecular spectra in subsection 3.2.
Finally,  subsection 3.3 will contain a discussion of the present results
and plans for further work.

\subsection{Formalism} 

By analogy to Eq. (2), $\Delta I=1$ staggering in nuclei can be measured
by the quantity
$$\Delta E_{1,\gamma}(I)= {1\over 16} (6 E_{1,\gamma}(I)-4 E_{1,\gamma}(I-1)
-4 E_{1,\gamma}(I+1) $$ 
\begin{equation}
+E_{1,\gamma}(I-2) + E_{1,\gamma}(I+2) ),
\end{equation} 
where
\begin{equation}
E_{1,\gamma}(I)= E(I+1)-E(I).
\end{equation} 
The transition energies $E_{1,\gamma} (I)$ are determined directly from
experiment.

In order to be able to use an expression similar to that of Eq. (12) for the
study of $\Delta J=1$ staggering in molecular bands we need transition
energies similar to those of Eq. (13), i.e. transition energies between levels
differing by one unit of angular momentum. However, Eqs (8) and (9)
can provide us only with transition energies between levels differing by
two units of angular momentum. In order to be able to determine the levels
with even $J$ from Eqs (8) or (9), one needs the bandhead energy $E(0)$.
Then one has
\begin{equation}
 E_{v_{upper}}(2) = E_{v_{upper}}(0) + E^R(1)-E^P(1),
\end{equation} 
\begin{equation}
 E_{v_{upper}}(4) = E_{v_{upper}}(2) + E^R(3)-E^P(3), \ldots
\end{equation} 
\begin{equation}
 E_{v_{lower}}(2)=  E_{v_{lower}}(0) + E^R(0)-E^P(2),
\end{equation} 
\begin{equation}
E_{v_{lower}}(4)=  E_{v_{lower}}(2) + E^R(2)-E^P(4), \ldots
\end{equation} 
In order to be able to determine the levels with odd $J$ from Eqs (8) and (9)
in an analogous way, one needs $E(1)$. Then
\begin{equation}
 E_{v_{upper}}(3)= E_{v_{upper}}(1)+E^R(2)-E^P(2),
\end{equation} 
\begin{equation}
 E_{v_{upper}}(5)= E_{v_{upper}}(3)+E^R(4)-E^P(4), \ldots
\end{equation} 
\begin{equation}
 E_{v_{lower}}(3)= E_{v_{lower}}(1)+E^R(1)-E^P(3),
\end{equation} 
\begin{equation}
 E_{v_{lower}}(5)= E_{v_{lower}}(3)+E^R(3)-E^P(5), \ldots
\end{equation} 

For the determination of $E(0)$ and
$E(1)$ one can use the overall fit of the
experimental data (for the R and P branches) by a Dunham expansion
\cite{Dunham}
\begin{equation}
E(J)= T_v + B_v J(J+1) -D_v [J(J+1)]^2 + H_v [J(J+1)]^3 + L_v [J(J+1)]^4,
\end{equation} 
which is usually given in the experimental papers.

After determining the energy levels by the procedure described above, we
estimate $\Delta J=1$ staggering by using the following analogue of  Eq. (12),
$$  \Delta E_{1,v} (J)= {1\over 16} (6 DE_{1,v}(J)-4 DE_{1,v}(J-1)
-4 DE_{1,v}(J+1) $$
\begin{equation}
+DE_{1,v}(J-2) +DE_{1,v}(J+2)),
\end{equation} 
where
\begin{equation}
DE_{1,v}(J)= E_v(J) - E_v(J-1).
\end{equation} 
Using Eq. (24) one can put Eq. (23) in the sometimes more convenient form
$$ \Delta E_{1,v}(J)= {1\over 16} (10 E_v(J)-10 E_v(J-1) + 5 E_v(J-2)
-5 E_v(J+1) $$
\begin{equation}
+E_v(J+2) -E_v(J-3)).
\end{equation} 

In realistic cases the first few values of $E^R(J)$ and $E^P(J)$ might
be experimentally unknown. In this case one is forced to determine the first
few values of $E(J)$ using the Dunham expansion of Eq. (22) and then
continue by using the Eqs (14)--(21) from the appropriate point on.
Denoting by $J_{io}$ the ``initial'' value of odd $J$, on which we
are building through the series of equations starting with Eqs (18)--(21)
the energy levels of odd $J$, and by $J_{ie}$ the ``initial'' value of
even $J$, on which we are building through the series of equations starting
with Eqs (14)--(17) the energy levels of even $J$, we find that the error
for the levels with odd $J$ is
\begin{equation}
Err(E(J)) = D(J_{io})+ (J-J_{io}) \epsilon,
\end{equation} 
while the error for the levels with even $J$ is
\begin{equation}
Err(E(J))= D(J_{ie})+(J-J_{ie}) \epsilon,
\end{equation} 
where $D(J_{io})$ and $D(J_{ie})$ are the uncertainties of the levels
$E(J_{io})$ and $E(J_{ie})$ respectively, which are determined through
the Dunham expansion of Eq. (22), while $\epsilon$ is the error
accompanying each $E^R(J)$ or $E^P(J)$ level, which in most
experimental works has a constant value for all levels.

Using Eqs (26) and (27) in Eq. (25) one easily sees that the uncertainty
of the $\Delta J=1$ staggering measure $\Delta E_{1,v}(J)$ is
\begin{equation}
Err(\Delta E_{1,v}(J)) = D(J_{io})+ D(J_{ie}) + (2J-J_{io}-J_{ie}-1) \epsilon.
\end{equation} 
This equation is valid for $J\geq {\rm max}\{ J_{io}, J_{ie} \}+3$.
For smaller values of $J$ one has to calculate the uncertainty
directly from Eq. (25).

\subsection{Analysis of experimental data} 

\subsubsection{YD} 

We have applied the formalism described above to the 0--1, 1--1, 1--2,
2--2 transitions of the C$^1\Sigma^+$--X$^1\Sigma^+$ system of YD \cite{YD}.
We have focused attention on the ground state  X$^1\Sigma^+$, which is known
to be free from $\Delta J=2$ staggering effects
(see subsection 2.2), while the
C$^1\Sigma^+$ state is known to exhibit $\Delta J=2$ staggering effects,
which are fingerprints of interband interactions (bandcrossings),
as we have seen in subsection 2.2.
Using the formalism of subsection 3.1, we calculated the
$\Delta J=1$ staggering measure $\Delta E_1(J)$ of Eq. (23)
for the $v=1$ band of the X$^1\Sigma^+$ state (Fig. 10a, 10b) and
for the $v=2$ band of the X$^1\Sigma^+$ state (Fig. 10c, 10d).
At this point the following comments are in place:

1) In all cases the levels $E(0)$, $E(1)$, $E(2)$, $E(3)$ have been determined
using the Dunham expansion of Eq. (22) and the Dunham coefficients given
in Table II of Ref. \cite{YD}. This has been done because $E^R(1)$ is
missing in the tables of the 1--1 and 2--2 transitions \cite{YD}, so
that Eq. (20) cannot be used for the determination of $E(3)$. In the cases
of the 0--1 and 1--2 transitions, $E^R(1)$ is known, but we prefered to
calculate $E(3)$ from the Dunham expansion in these cases as well,
in order to treat the pairs of cases 0--1, 1--1 and 1--2, 2--2 on equal
footing, since we intend to make comparisons between them.

2) For the calculation of errors we have taken into account the errors
of the Dunham coefficients given in Table II of Ref. \cite{YD}, as well
as the fact that the accuracy of the members of the R- and P-branches
is $\epsilon=\pm 0.002$ cm$^{-1}$ \cite{YD}. It is clear that
the large size of the error bars is due to the accumulation
of errors caused by Eqs (14)--(21), as seen in Eqs (25)--(28).

3) In Figs 10a and 10b the $\Delta J=1$ staggering measure $\Delta E_1(J)$
for the $v=1$ band of the X$^1\Sigma^+$ state of YD is shown, calculated
from two different sources, the 0--1 and 1--1 transitions.
If a real $\Delta J=1$ staggering effect were
present, the two figures should have been identical, or at least consistent
with each other. However, they are completely different (even the maxima
and the minima appear at different values of $J$ in each figure), indicating
that what is seen is not a real physical effect, but random experimental
errors (buried in the large error bars, anyway).

4) Exactly the same comments as in 3) apply to Figs 10c and 10d,
where the $\Delta J=1$ staggering measure for the $v=2$ band of the
X$^1\Sigma^+$ state of YD is shown, calculated from two different sources,
the 1--2 and 2--2 transitions.

We conclude therefore that no $\Delta J=1$ staggering effect appears
in the $v=1$ and $v=2$ bands of the X$^1\Sigma^+$ state of YD, which are
free from $\Delta J=2$ staggering, as proved in subsection 2.2.

This negative result has the following physical implications. It is known
in nuclear spectroscopy that reflection asymmetric (pear-like) shapes
give rise to octupole bands, in which the positive parity states
($I^\pi=0^+$, $2^+$, $4^+$, \dots) are displaced reletively to
the negative parity states ($I_\pi=1^-$, $3^-$, $5^-$, \dots)
\cite{Phill,Schuler,Ahmad,Butler,Leander,LeanderSh,Krappe}.
Since a diatomic molecule consisting of two different atoms possesses
the same reflection asymmetry, one might think that $\Delta J=1$
staggering might be present in the rotational bands of such molecules.
Then YD, because of its large mass asymmetry, is a good testing ground
for this effect. The negative result obtained above can, however, be readily
explained. Nuclei with octupole deformation are supposed to be described by
double well potentials, the relative displacement of the negative parity
levels and the positive parity levels being attributed to the tunneling
through the barrier separating the wells
\cite{Leander,LeanderSh,Krappe}. (The relative displacement
vanishes in the limit in which the barrier separating the two wells
becomes infinitely high.) In the case of diatomic molecules the relevant
potential is well known \cite{Herz} to consist of a single well.
Therefore no tunneling effect is possible and, as a result, no relative
displacement of the positive parity levels and the negative parity levels
is seen.

\subsubsection{CrD} 

The formalism of subsection 3.1 has in addition been applied to a more
complicated case, the one of the 0--0 and 1--0 transitions of the
A$^6\Sigma^+$--X$^6\Sigma^+$ system of CrD \cite{CrD}.
We have focused our attention on the ground state X$^6 \Sigma^+$, which is
known to be free from $\Delta N=2$ staggering effects (see subsection 2.2),
while the A$^6 \Sigma^+$ state is known from subsection 2.2
to exhibit $\Delta N=2$ staggering effects, which are
fingerprints of interband interactions (bandcrossings).
The CrD system considered here has several differences from the YD system
considered in the previous subsection, which are briefly listed here:

1) The present system of CrD involves sextet electronic states. As a result,
each band of the A$^6\Sigma^+$--X$^6\Sigma^+$ transition consists of
six R- and six P-branches, labelled as R1, R2, \dots, R6 and P1, P2, \dots, P6
respectively \cite{CrD}. In the present study we use the R3 and P3 branches,
but similar results are obtained for the other branches as well.

2) Because of the presence of spin--rotation interactions and spin--spin
interactions, the energy levels cannot be fitted by a Dunham expansion
in terms of the total angular momentum $J$, but by a more complicated
Hamiltonian, the $N^2$ Hamiltonian for a $^6\Sigma$ state \cite{Brown,Gordon}.
This Hamiltonian, in addition to a Dunham expansion in terms of $N$
(the rotational angular momentum, which in this case is different
from the total angular momentum ${\bf J}={\bf N}+{\bf S}$, where $S$
the spin), contains terms describing the spin--rotation interactions
(preceded by three $\gamma$ coefficients), as well as terms describing the
spin--spin interactions (preceded by two $\lambda$ coefficients
\cite{CrD,Brown}).

In the present study we have calculated the staggering measure
of Eq. (23) for the $v=0$ band of the X$^6\Sigma^+$ state of CrD, using
the R3 and P3 branches of the 0--0 (Fig. 11a) and 1--0 (Fig. 11b)
transitions of the
A$^6\Sigma^+$--X$^6\Sigma^+$ system. Since in this case the Dunham
expansion involves the rotational angular momentum $N$, and not the
total angular momentum $J$, the formalism of subsection 3.1 has been used
with $J$ replaced by $N$ everywhere. This is why the calculated
staggering measure of Eq. (23) is in this case denoted by $\Delta E_1(N)$
and not by $\Delta E_1(J)$, the relevant effect being called $\Delta N=1$
staggering instead of $\Delta J=1$ staggering.
At this point the following comments are in place:

1) In both cases the levels $E(0)$, $E(1)$, $E(2)$, $E(3)$, $E(4)$ have
been determined using the Dunham expansion of Eq. (22) (with $J$ replaced
by $N$) and the Dunham coefficients given in Table V of Ref. \cite{CrD}.
This has been done because $E^R(2)$ is missing in the tables of the 0--0
transitions \cite{CrD}, so that Eq. (17) cannot be used for the determination
of $E(4)$. In the case of the 1--0 transitions, $E^R(2)$ is known, but
we prefered to calculate $E(4)$ from the Dunham expansion in this
case as well, in order to treat the cases 0--0 and 1--0 on equal footing,
since we intend to make comparisons between them.

2) For the calculation of errors we have taken into account the errors
of the Dunham coefficients given in Table V of Ref. \cite{CrD}, as well
as the fact that the accuracy of the members of the R- and P- branches
is $\epsilon=\pm 0.001$ cm$^{-1}$ for the 0--0 transitions and
$\epsilon=\pm 0.003$ cm$^{-1}$ for the 1--0 transitions \cite{CrD}.
In this case it is clear,
as in the previous one, that the large size of the error bars is due to
the accumulation of errors caused by Eqs (14)--(21), as seen in Eqs
(25)--(28).

3) In Figs 11a and 11b the $\Delta N=1$ staggering measure $\Delta E_1(N)$
for the $v=0$ band of the X$^6\Sigma^+$ state of CrD is shown, calculated
from two different sources, the 0--0 and 1--0 transitions. The two
figures are nearly identical. The maxima and the minima appear at the
same values of $N$ in both figures, while even the amplitude of the effect
is almost the same in both figures. It should be pointed out, however,
that the error bars in Fig. 11b have been made smaller by a factor of three,
in order to the accommodated in the figure.

We conclude therefore that in the $v=0$ band of the X$^6\Sigma^+$ state of
CrD the two different calculations give consistent results, despite the
error accumulation mentioned above. The result looks like $\Delta N=1$
staggering of almost constant amplitude. The reason behind the appearance
of this staggering is, however clear: It is due to the omission of the
spin-rotation and spin-spin terms  of the $N^2$ Hamiltonian mentioned
above \cite{CrD,Brown,Gordon}.
As a result, we have not discovered any new physical effect.
What we have demonstated, is that Eq. (23) is a very sensitive probe,
which can uncover small deviations from the pure rotational behaviour.
However, special care should be taken when using it, because of the
accumulation of errors, which is inherent in this method. This problem
is avoided by producing results for the same band from two different
sets of data, as done above. If both sets lead to consistent results,
some effect is present. If the two sets give randomly different results,
it is clear that no effect is present.

It should be pointed out at this point that the appearance of $\Delta J=1$
staggering (or $\Delta N=1$ staggering) does {\sl not} mean that an effect
with oscillatory behaviour is present. Indeed, suppose that the energy
levels of a band follow the $E(J)=A J(J+1)$ rule, but to the odd levels
a constant term $c$ is added. It is then clear from Eq. (25) that we are
going to obtain $\Delta E_1(J)= +c$ for odd values of $J$, and
$\Delta E_1(J)=-c$ for even values of $J$, obtaining in this way perfect
$\Delta J=1$ staggering of constant amplitude $c$, without the presence
of any oscillatory effect. This comment directly applies to the results
presented in Fig. 11. The presence of $\Delta N=1$ staggering of almost
constant amplitude is
essentially due to the omission of the rotation--spin and spin--spin
interactions in the calculation of the $E(3)$ and $E(4)$ levels.
The difference of the omitted terms in the $N=3$ and $N=4$ cases plays
the role of $c$ in Fig. 11.

\subsection{Discussion} 

In this section we have addressed the question of the possible
existence of $\Delta J=1$ staggering (i.e. of a relative displacement
of the odd levels with respect to the even levels)
in rotational bands of diatomic molecules, which are free from $\Delta J=2$
staggering (i.e. free from interband interactions/bandcrossings).
The main conclusions drawn are:

1) The YD bands studied indicate that there is no $\Delta J=1$ staggering,
which could be due to the mass asymmetry of this molecule.

2) The CrD bands studied indicate that there is $\Delta N=1$ staggering,
which is, however, due to the spin--rotation and spin--spin interactions
present in the relevant states.

3) Based on the above results, we see that $\Delta J=1$ staggering is a
sensitive probe of deviations from the pure rotational behaviour.
Since the method of its calculation from the experimental data leads,
however, to error accumulation, one should always calculate the $\Delta J=1$
staggering measure for the same band from two different sets of data
and check the consistency of the results, absence of consistency meaning
absence of any real effect.

It is desirable to corroborate the above conclusions by studying
rotational bands of several additional molecules.

\section{Conclusions}

In this work we have examined if the effects of $\Delta J=2$ staggering
and $\Delta J=1$ staggering, which appear in nuclear spectroscopy,
appear also in rotational bands of diatomic molecules. For the $\Delta J=2$
staggering it has been found that it appears in certain electronically excited
rotational bands of diatomic molecules (YD, CrD, CrH, CoH), in which
it is attributed to interband interactions (bandcrossings). The
$\Delta J=1$ staggering has been examined in rotational bands free from
$\Delta J=2$ staggering, i.e. free from interband interactions
(bandcrossings). Bands of YD offer evidence for the absence of any $\Delta
J=1$ staggering effect due to the disparity of nuclear masses, while
bands of sextet electronic states of CrD demonstrate that $\Delta J=1$
staggering is a sensitive probe of deviations from rotational behaviour,
due in this particular case to the spin-rotation and spin-spin interactions.
We conclude therefore that both $\Delta J=2$ staggering and $\Delta J=1$
staggering are sensitive probes of perturbations in rotational bands
of diatomic molecules and do not constitute any new physical effect.

The number of rotational bands of diatomic molecules examined in the case
of the $\Delta J=2$ staggering is satisfactory. For the case of the
$\Delta J=1$ staggering it is desirable to corroborate the findings of the
present work through the examination of rotational bands of more diatomic
molecules.

\bigskip
{\bf Acknowledgements}
\medskip

One of the authors (PPR) acknowledges support from the Bulgarian Ministry
of Science and Education under contract $\Phi$-547.
Another author (NM) has been supported by the Bulgarian National Fund
for Scientific Research under contract no MU-F-02/98.
Three authors (DB,CD,NK) have been supported by the Greek Secretariat
of Research and Technology under contract PENED 95/1981.



\centerline{\bf Figure captions} 
\ms 

\ref {\bf  Fig. 1} 
$\Delta E_2(I)$ (in keV), calculated  from Eq. (2),
versus the angular frequency $\omega$ (in MeV), calculated from Eq. (7),
for various superdeformed bands in the nucleus $^{149}$Gd [13].
a) Band (a) of Ref. [13]. b) Band (d) of Ref. [13].

\ref {\bf  Fig. 2} 
$\Delta E_2(I)$ (in keV), calculated from Eq. (2),
versus the angular frequency $\omega$ (in MeV), calculated from Eq. (7),
for various superdeformed bands in the nucleus $^{194}$Hg [14].
a) Band 1 of Ref. [14]. b) Band 2 of Ref. [14].

\ref {\bf Fig. 3} 
$\Delta E_2(J)$ (in cm$^{-1}$), calculated from Eq. (11),
for various bands of the YD molecule [34].
a) Odd levels of the $v=1$ C$^1\Sigma^+$ band calculated from the
data of the 1--1 C$^1\Sigma^+$--X$^1\Sigma^+$ transitions.
b) Even levels of the previous band.
c) Odd levels of the $v=1$ C$^1\Sigma^+$ band calculated from the 1--2
C$^1\Sigma^+$--X$^1\Sigma^+$ transitions.
d) Even levels of the previous band.

\ref {\bf Fig. 4} 
$\Delta E_2(J)$ (in cm$^{-1}$), calculated from Eq. (11),
for various bands of the YD molecule [34].
a) Odd levels of the $v=2$ C$^1\Sigma^+$ band calculated from the
data of the 2--2 C$^1\Sigma^+$--X$^1\Sigma^+$ transitions.
b) Even levels of the previous band.
c) Odd levels of the $v=2$ C$^1\Sigma^+$ band calculated from the 2--3
C$^1\Sigma^+$--X$^1\Sigma^+$ transitions.
d) Even levels of the previous band.
The experimental error in all cases is $\pm 0.006$ cm$^{-1}$ and therefore
is hardly or not seen.

\ref {\bf Fig. 5} 
$\Delta E_2(J)$ (in cm$^{-1}$), calculated from Eq. (11),
for various bands of the YD molecule [34]. 
a) Odd levels of the $v=1$ X$^1\Sigma^+$ band calculated from the
data of the 1--1 C$^1\Sigma^+$--X$^1\Sigma^+$ transitions.
b) Even levels of the previous band.

\ref {\bf Fig. 6}
$\Delta E_2(J)$ (in cm$^{-1}$), calculated from Eq. (11),
for various bands of the YD molecule [34]. 
a) Odd levels of the $v=2$ X$^1\Sigma^+$ band calculated from the
data of the 1--2 C$^1\Sigma^+$--X$^1\Sigma^+$ transitions.
b) Even levels of the previous band.
c) Odd levels of the $v=2$ X$^1\Sigma^+$ band calculated from the 2--2
C$^1\Sigma^+$--X$^1\Sigma^+$ transitions.
d) Even levels of the previous band.

\ref {\bf Fig. 7}
$\Delta E_2(N)$ (in cm$^{-1}$), calculated from Eq. (11),
for various bands of the CrD molecule [35]. 
a) Odd levels of the $v=0$ A$^6\Sigma^+$ band calculated from the
data (R2, P2 branches) of the 0--0 A$^6\Sigma^+$--X$^6\Sigma^+$ transitions.
b) Even levels of the previous band.
The experimental error in all cases is $\pm 0.006$ cm$^{-1}$
and therefore is not seen.

\ref {\bf Fig. 8}
$\Delta E_2(N)$ (in cm$^{-1}$), calculated from Eq. (11),
for various bands of the CrH molecule [36]. 
a) Odd levels of the $v=0$ A$^6\Sigma^+$ band calculated from the
data (R2, P2 branches)
of the 0--0 A$^6\Sigma^+$--X$^6\Sigma^+$ transitions.
b) Even levels of the previous band.
The experimental error in all cases is $\pm 0.004$ cm$^{-1}$
and therefore is not seen.

\ref {\bf Fig. 9} 
$\Delta E_2(J)$ (in cm$^{-1}$), calculated from Eq. (11),
for various bands of the CoH molecule [37]. 
a) Odd levels of the $v=0$ A$'$$^{3}\Phi_4$ band calculated from the
data (Ree, Pee branches) of the 0--1 A$'$$^{3}\Phi_4$--X$^3\Phi_4$
transitions.
b) Even levels of the previous band.
The experimental error in all cases is $\pm 0.01$ cm$^{-1}$
and therefore is not seen.

\ref {\bf Fig. 10} 
$\Delta E_1(J)$ (in cm$^{-1}$), calculated from Eq. (23),
for various bands of the YD molecule [34].
a) Levels of the $v=1$ X$^1\Sigma^+$ band calculated from the
data of the 0--1 C$^1\Sigma^+$--X$^1\Sigma^+$ transitions.
b) Levels of the $v=1$ X$^1\Sigma^+$ band calculated from the data of the 1--1
C$^1\Sigma^+$--X$^1\Sigma^+$ transitions.
c) Levels of the $v=2$ X$^1\Sigma^+$ band calculated from the data of the
1--2 C$^1\Sigma^+$--X$^1\Sigma^+$ transitions.
d) Levels of the $v=2$ X$^1\Sigma^+$ band calculated from the data of the
2--2 C$^1\Sigma^+$--X$^1\Sigma^+$ transitions.

\ref {\bf Fig. 11}
$\Delta E_1(N)$ (in cm$^{-1}$), calculated from Eq. (23),
for various bands of the CrD molecule [35].
a) Levels of the $v=0$ X$^6\Sigma^+$ band calculated from the data of the
0--0 A$^6\Sigma^+$--X$^6\Sigma^+$ transitions (R3, P3 branches).
b) Levels of the $v=0$ X$^6\Sigma^+$ band calculated from the data of the
1--0 A$^6\Sigma^+$--X$^6\Sigma^+$ transitions (R3, P3 branches).
The error bars in case (b) have been divided by a factor of 3, in order
to be accommodated within the figure.

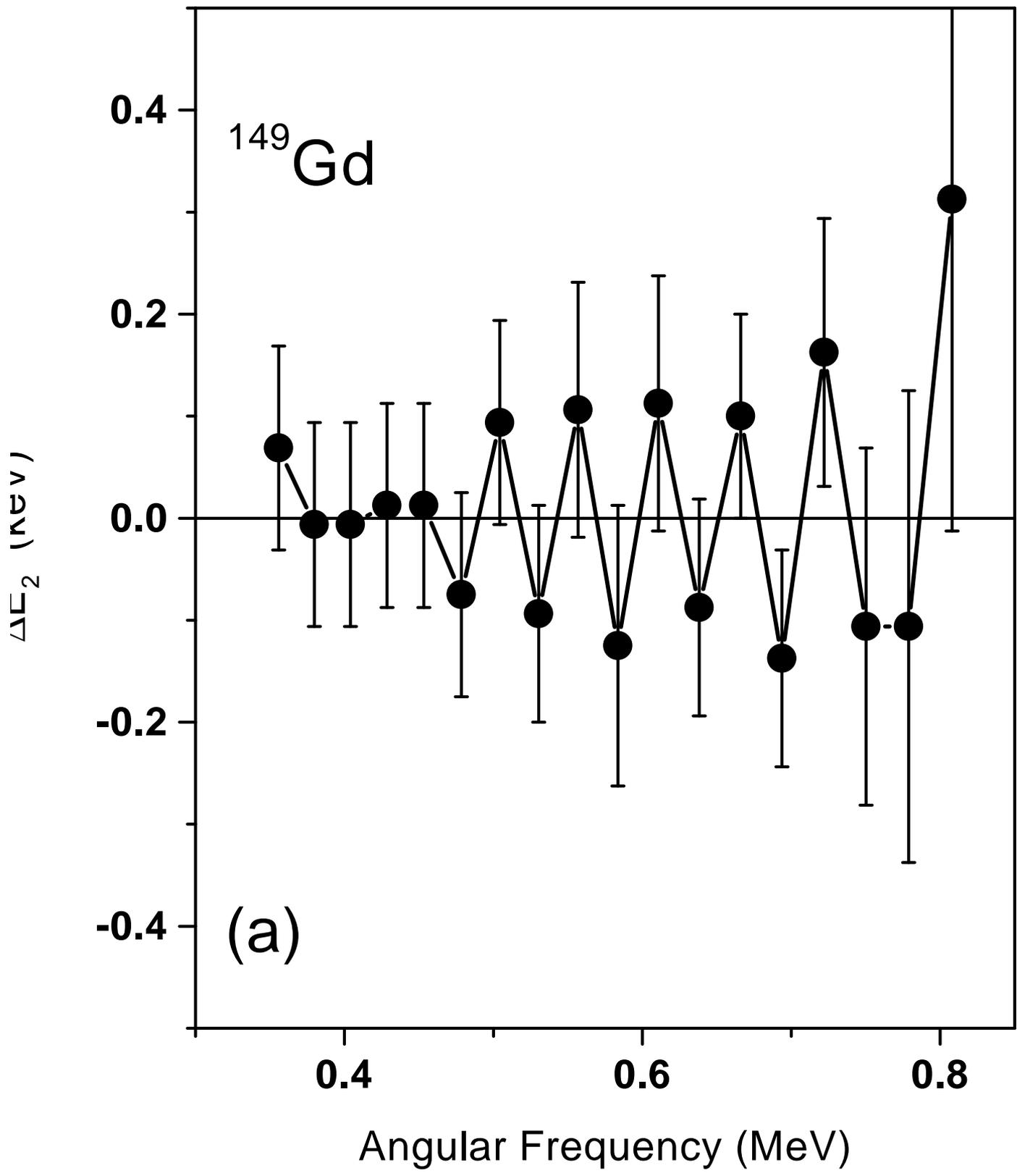
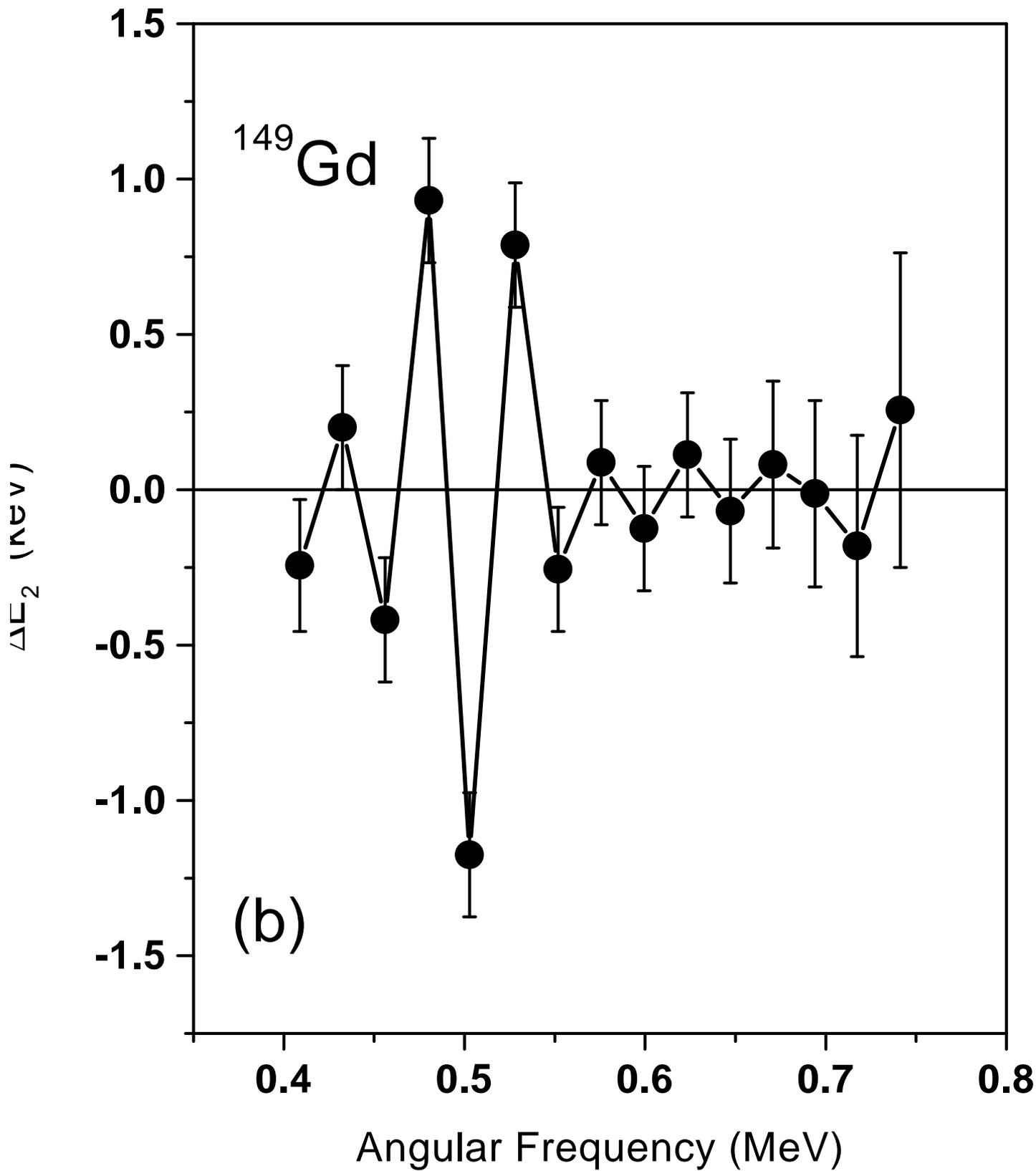
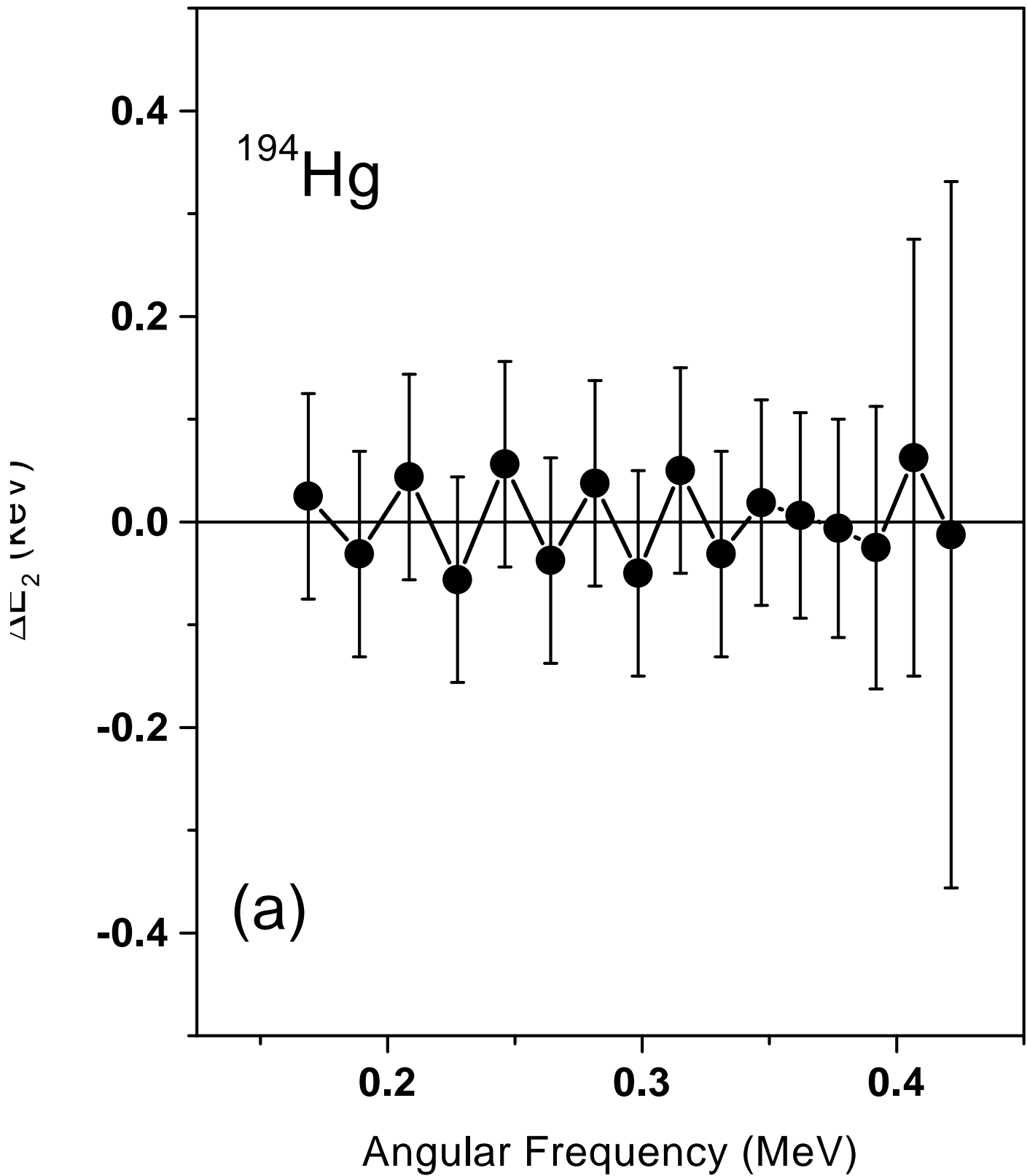
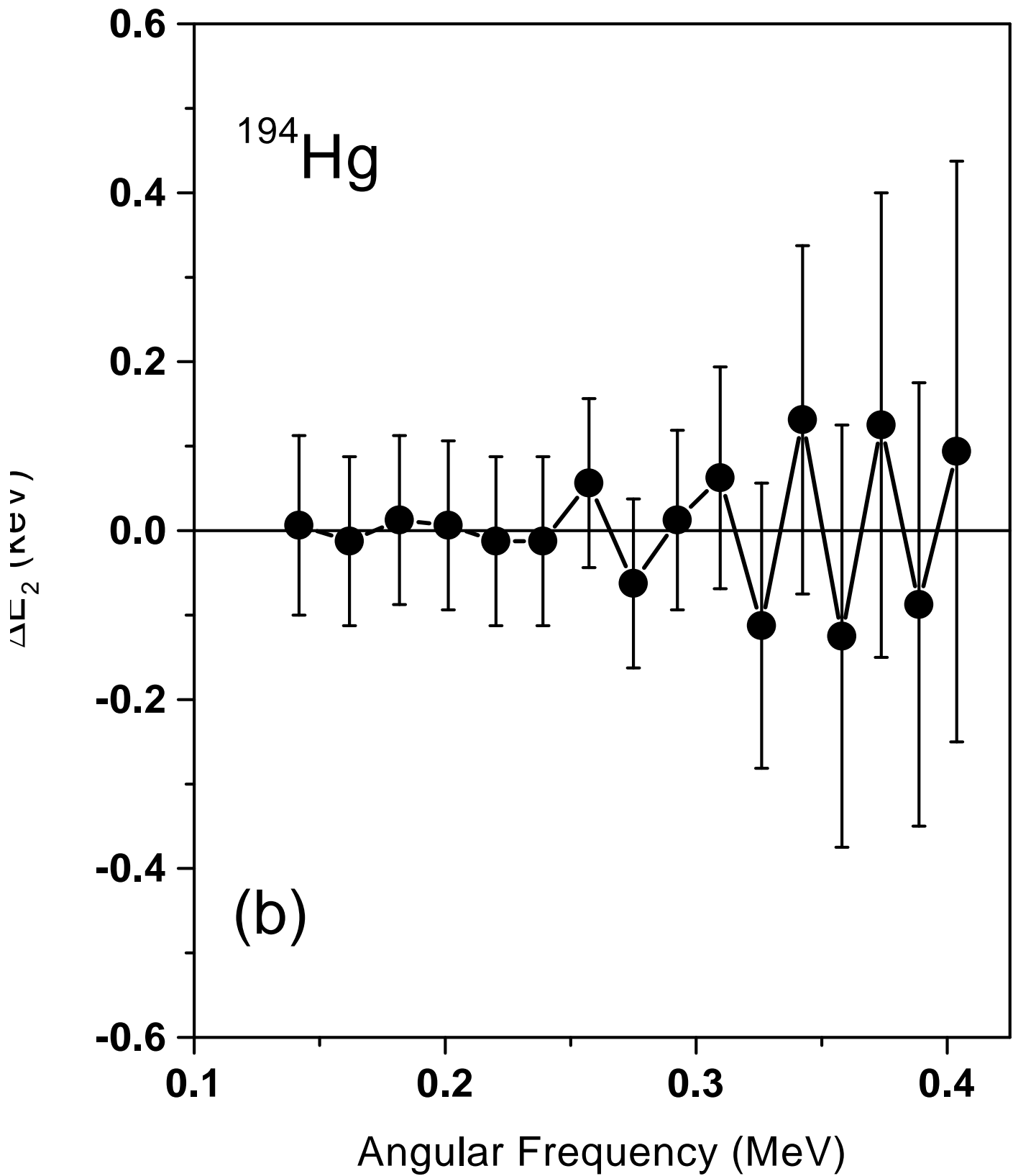
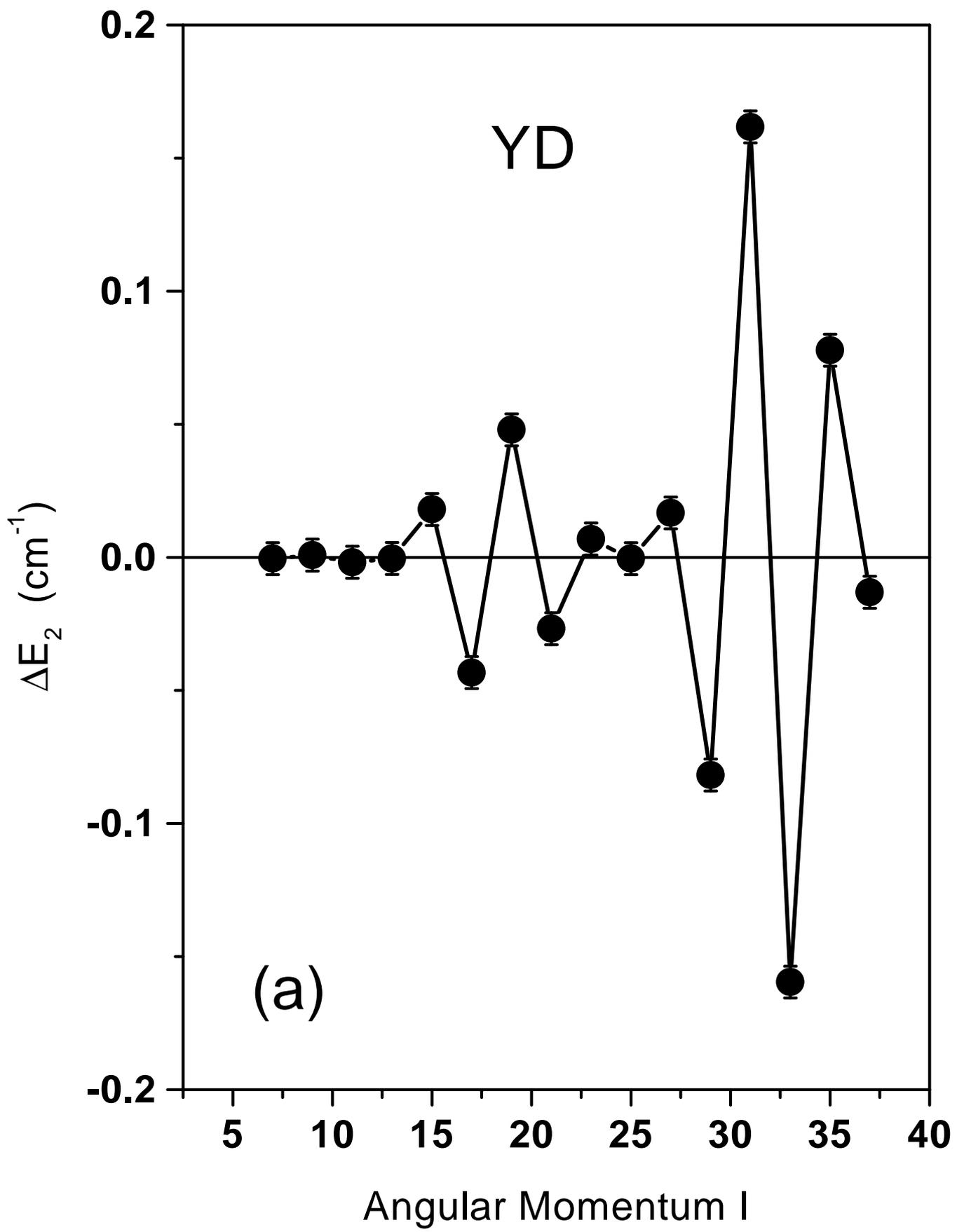
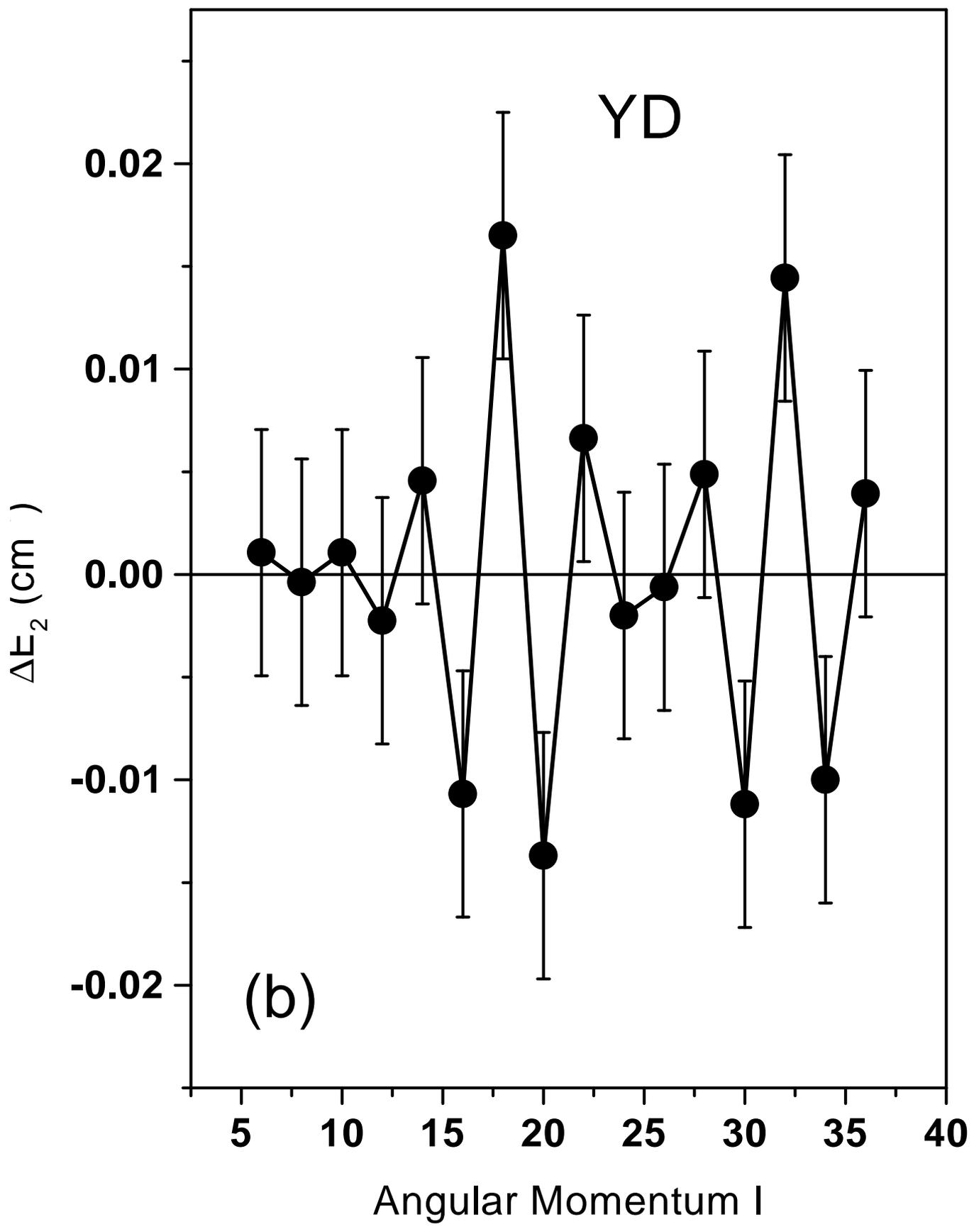
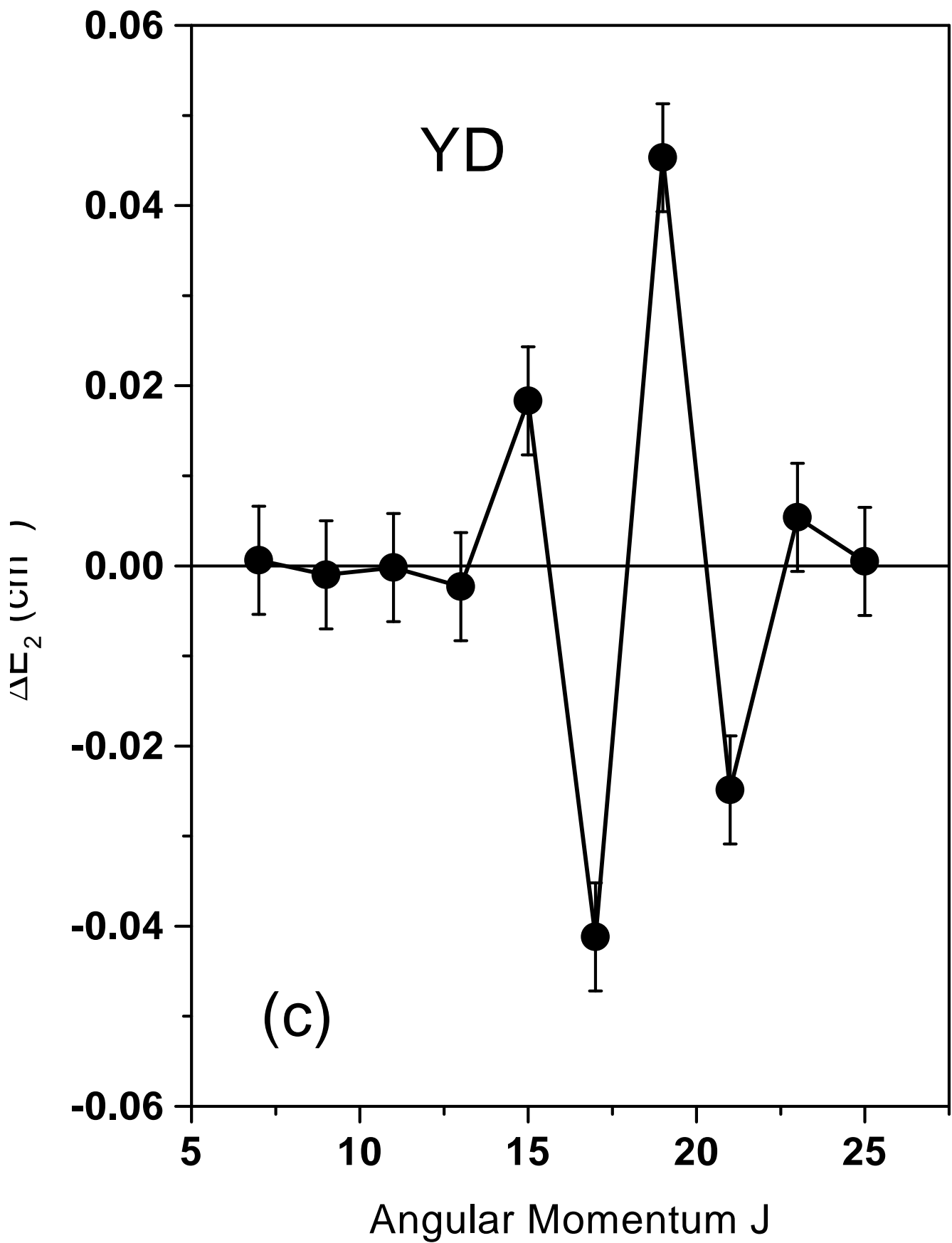
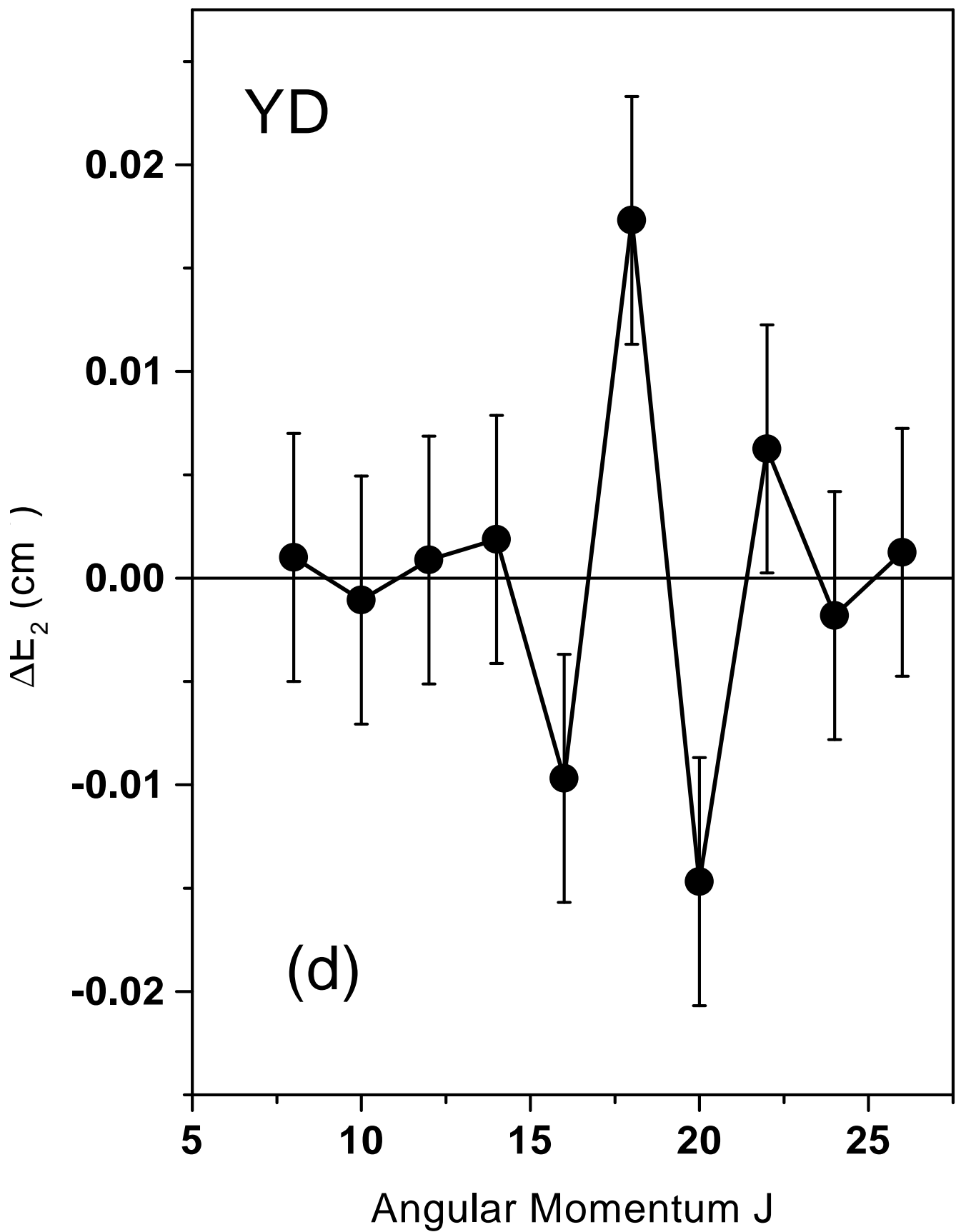
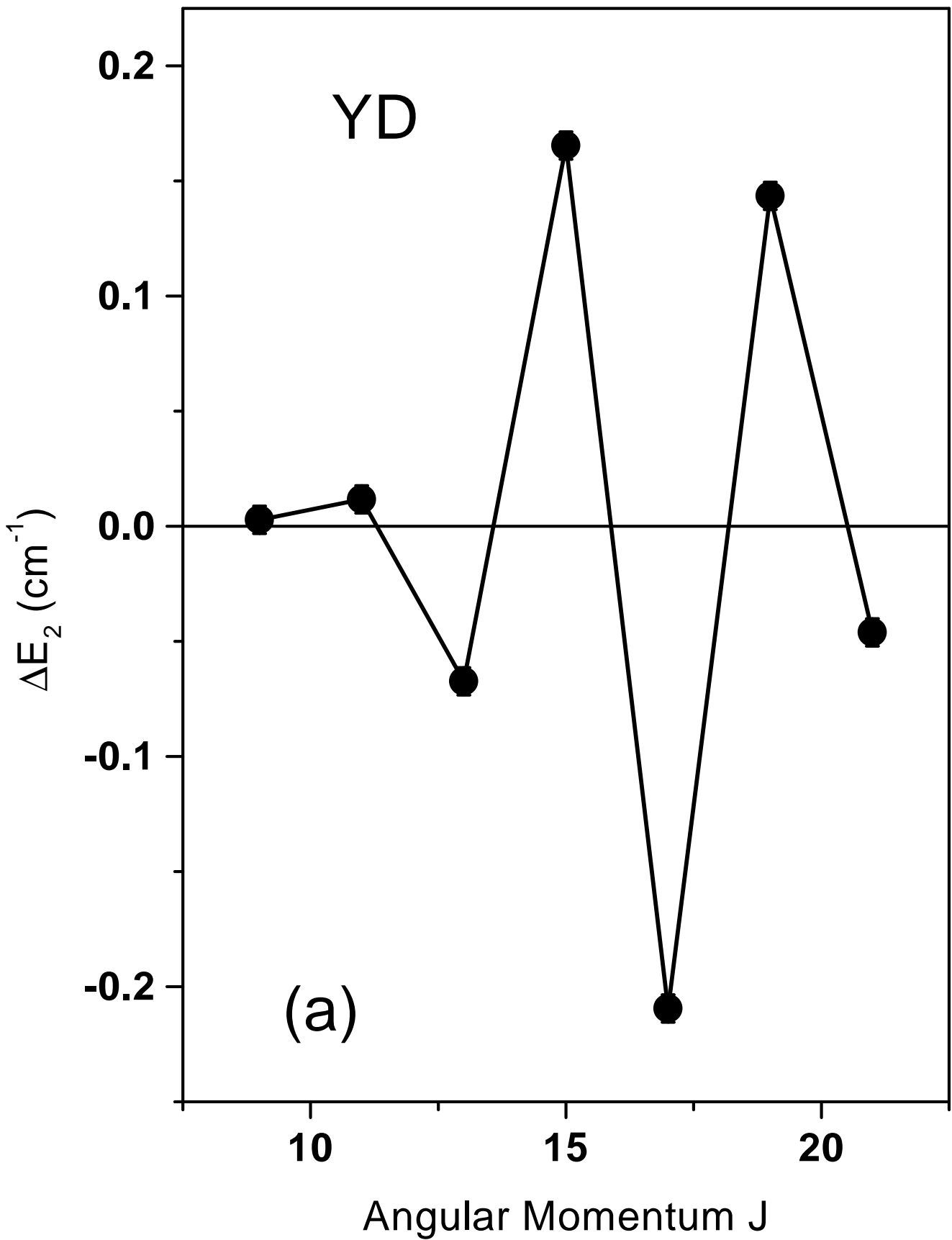
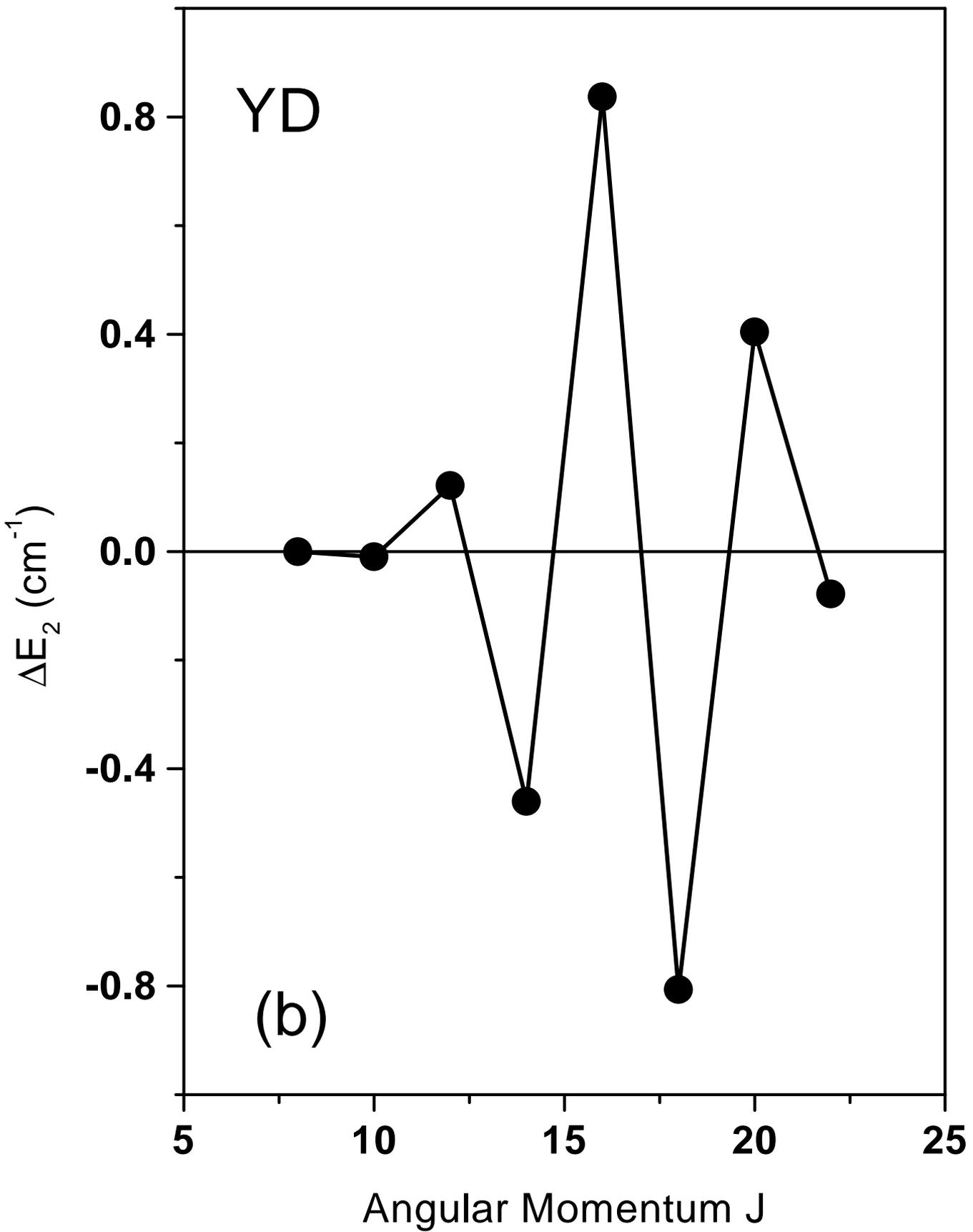
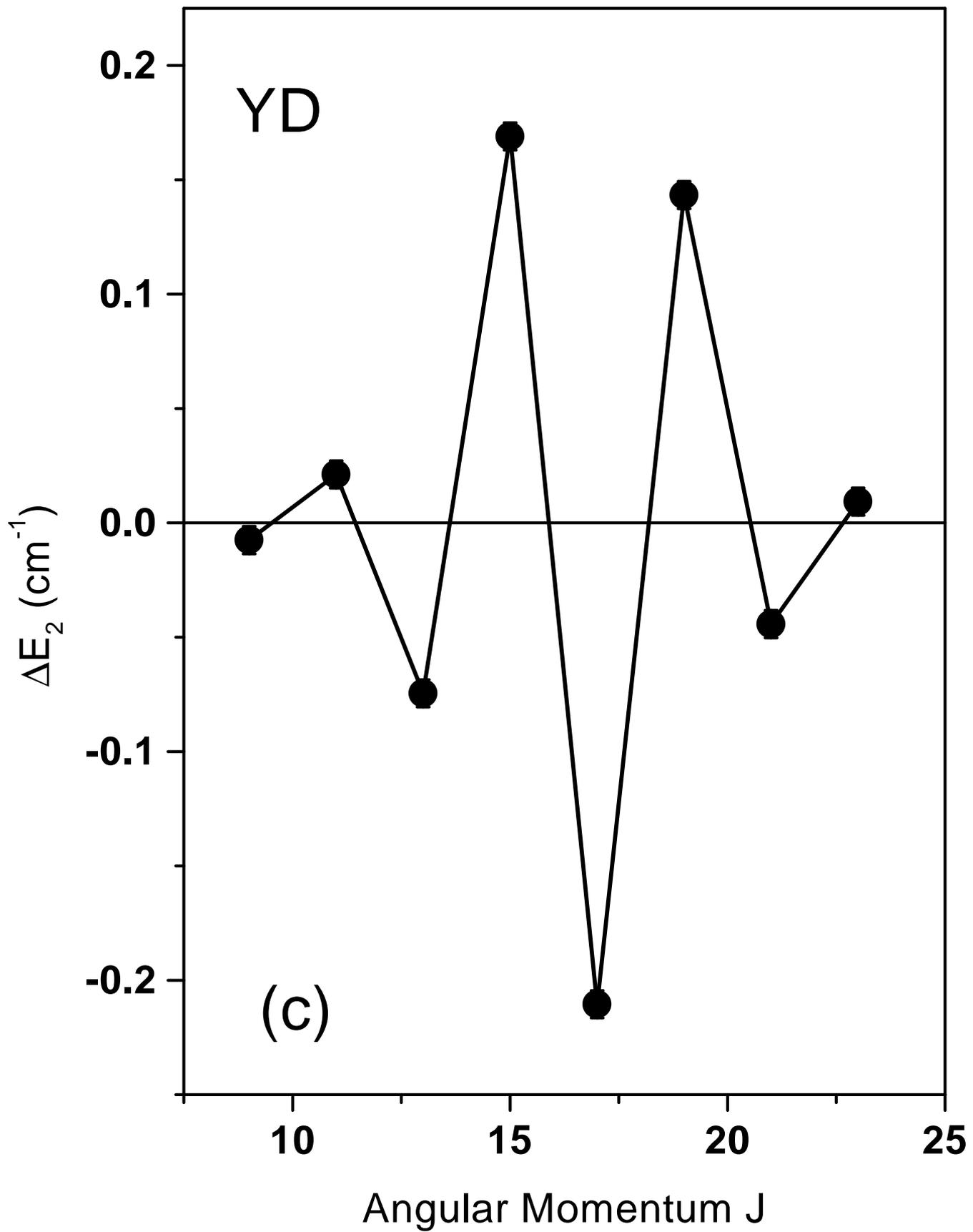
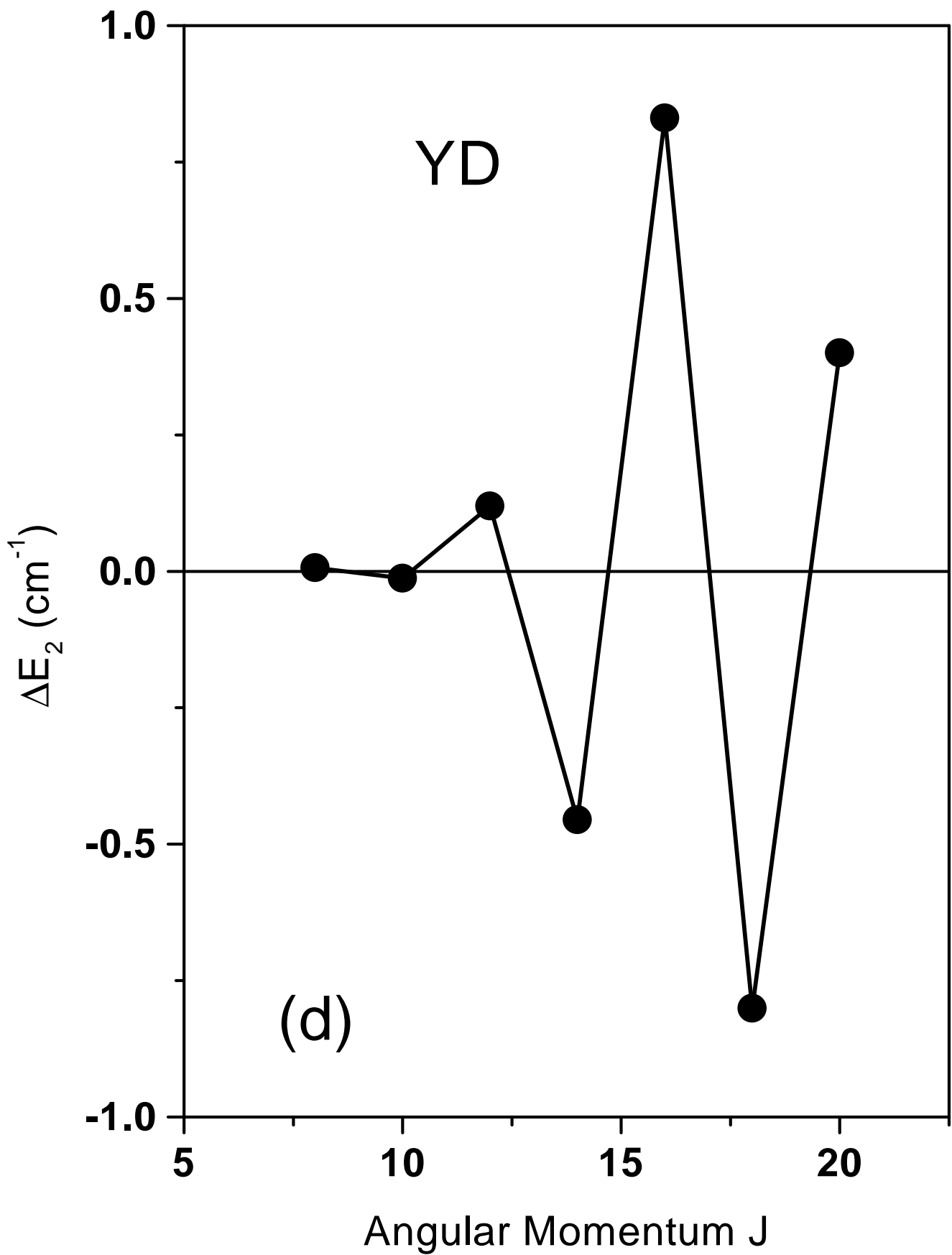
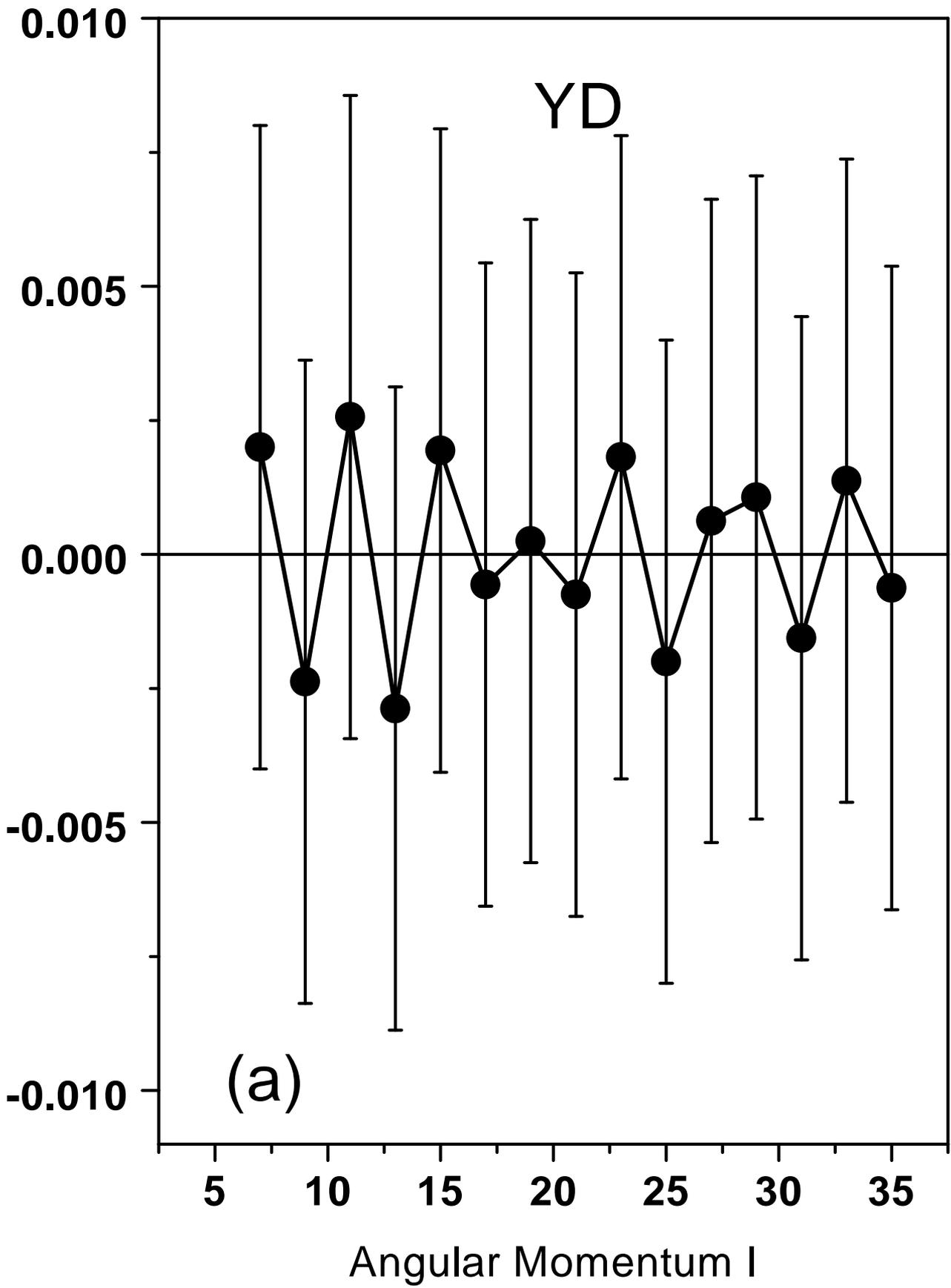
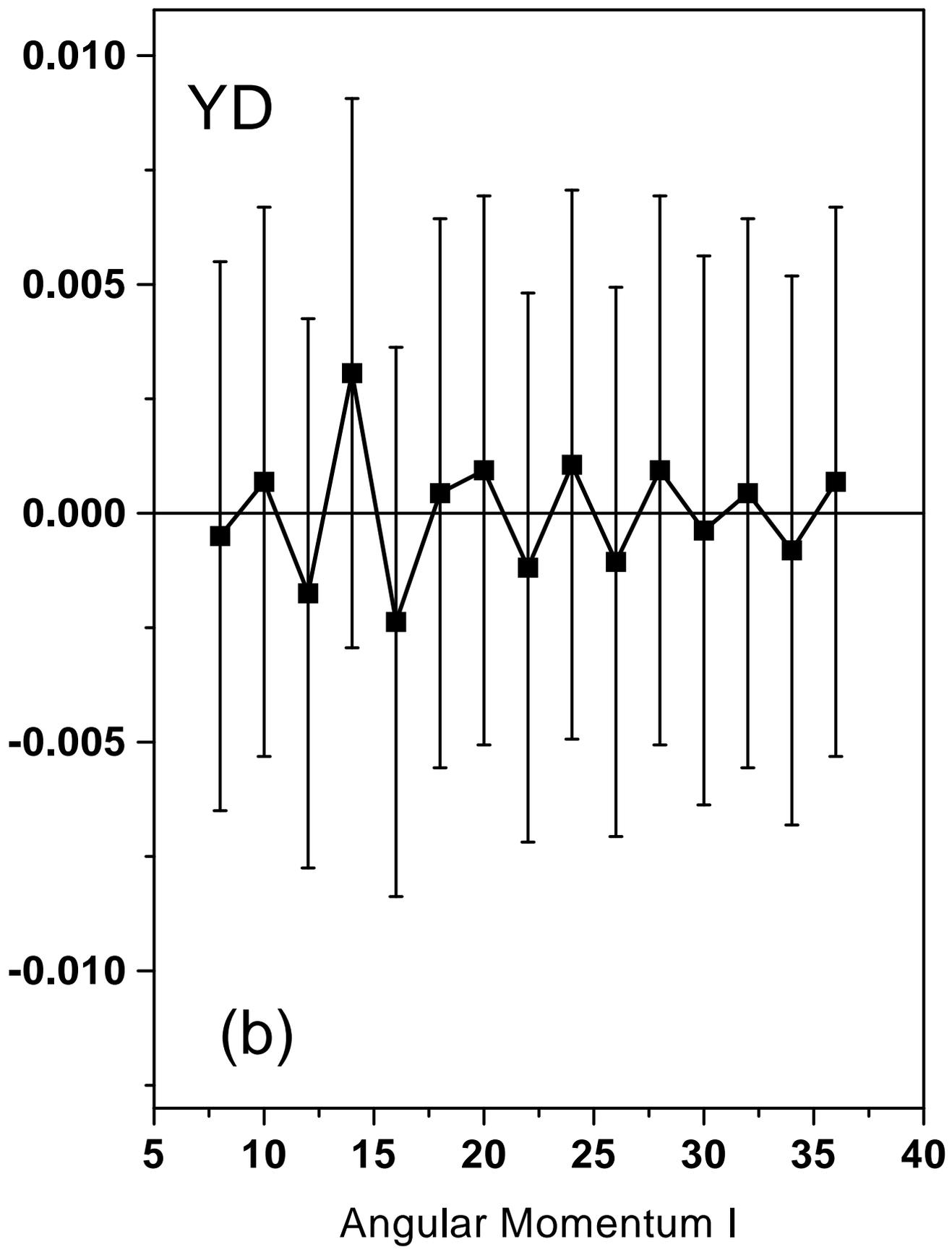
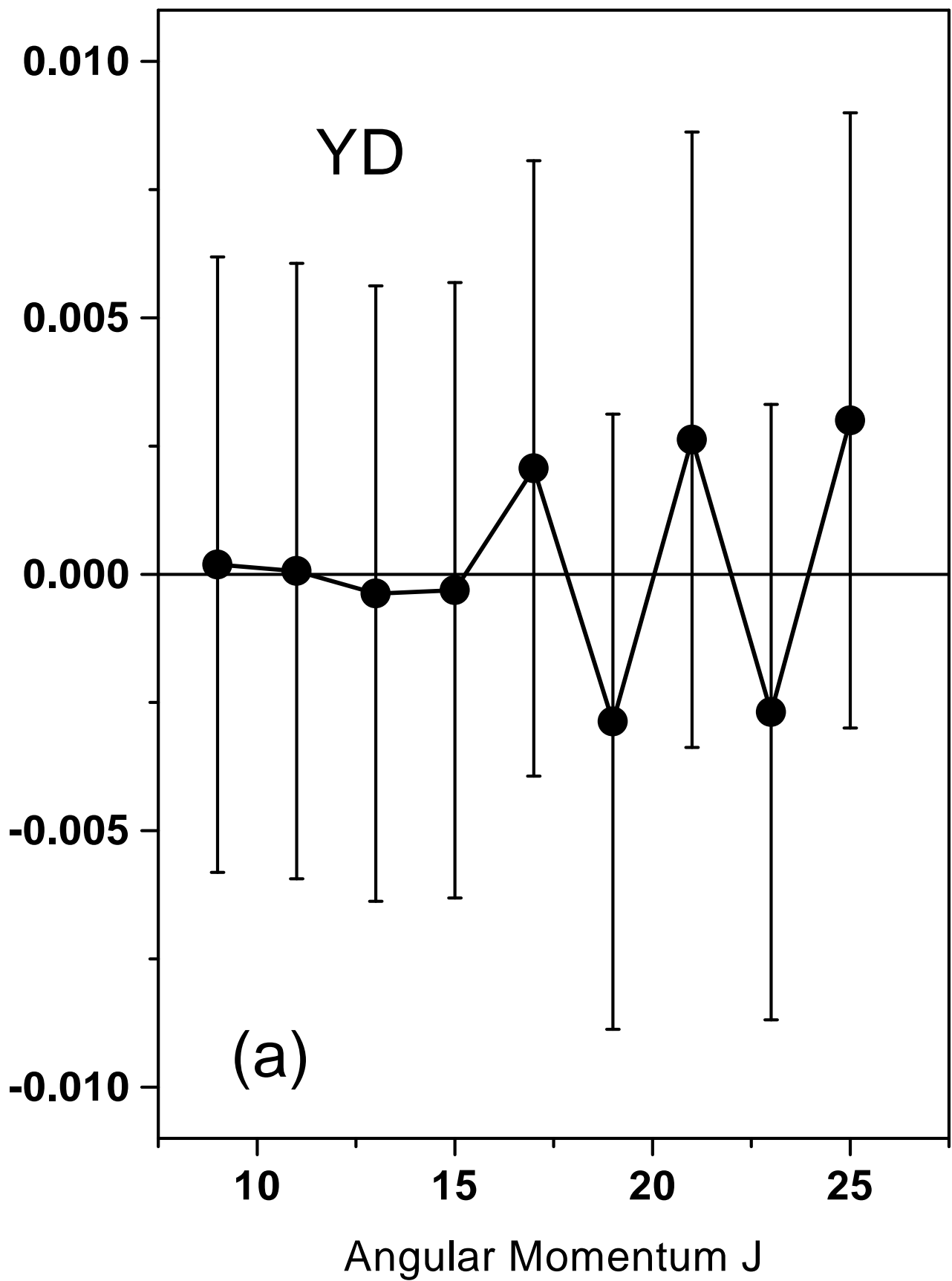
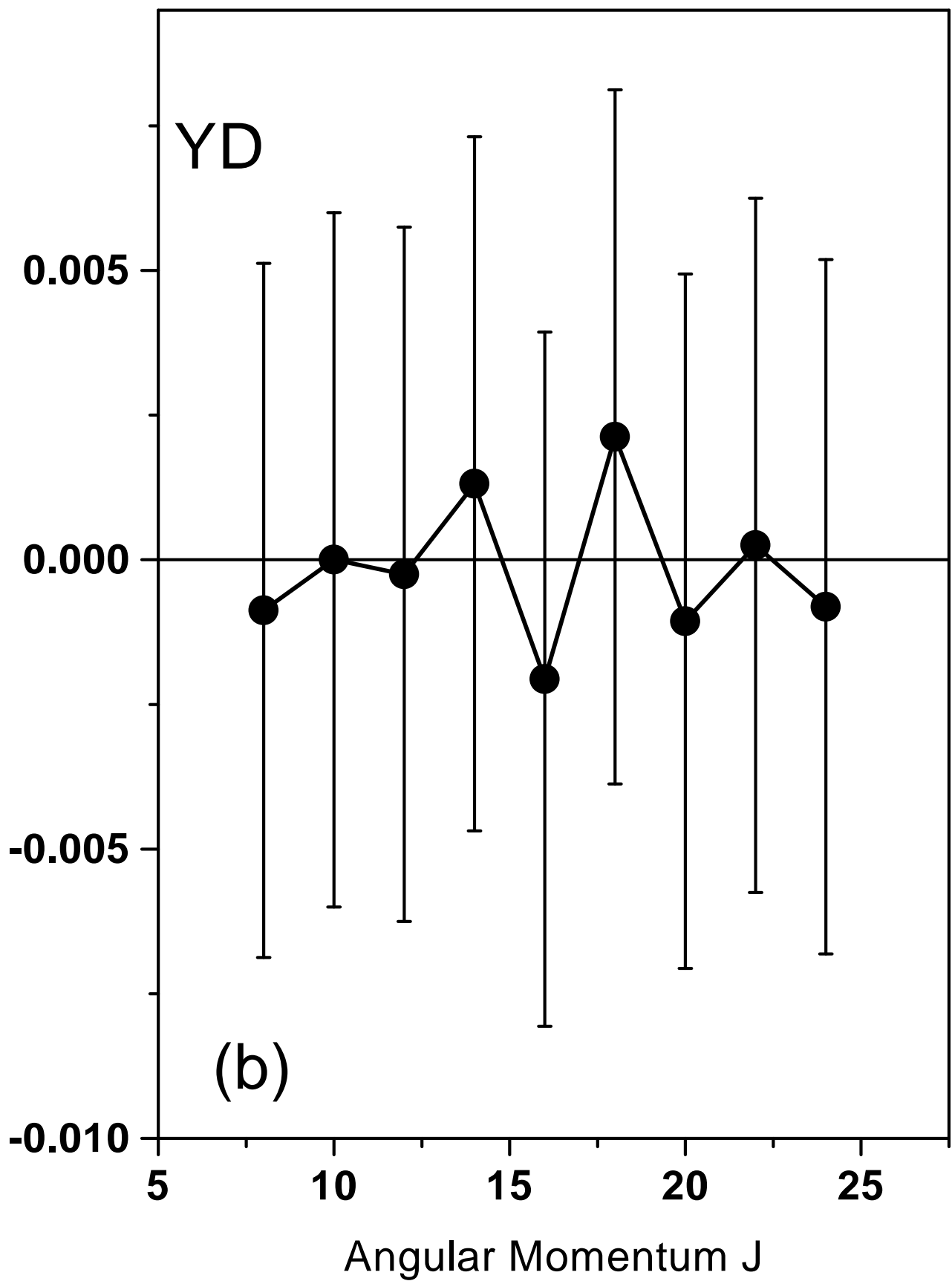
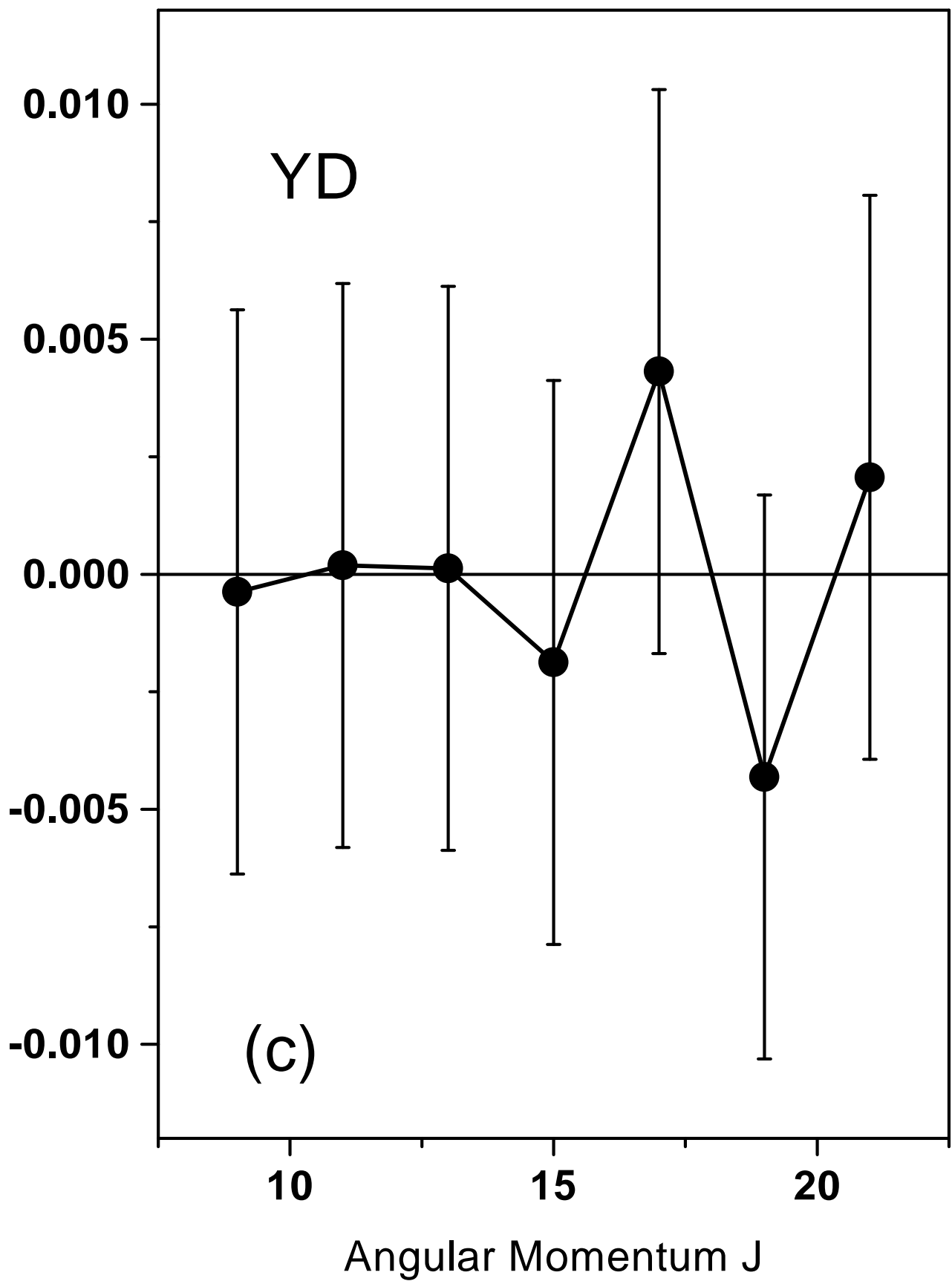
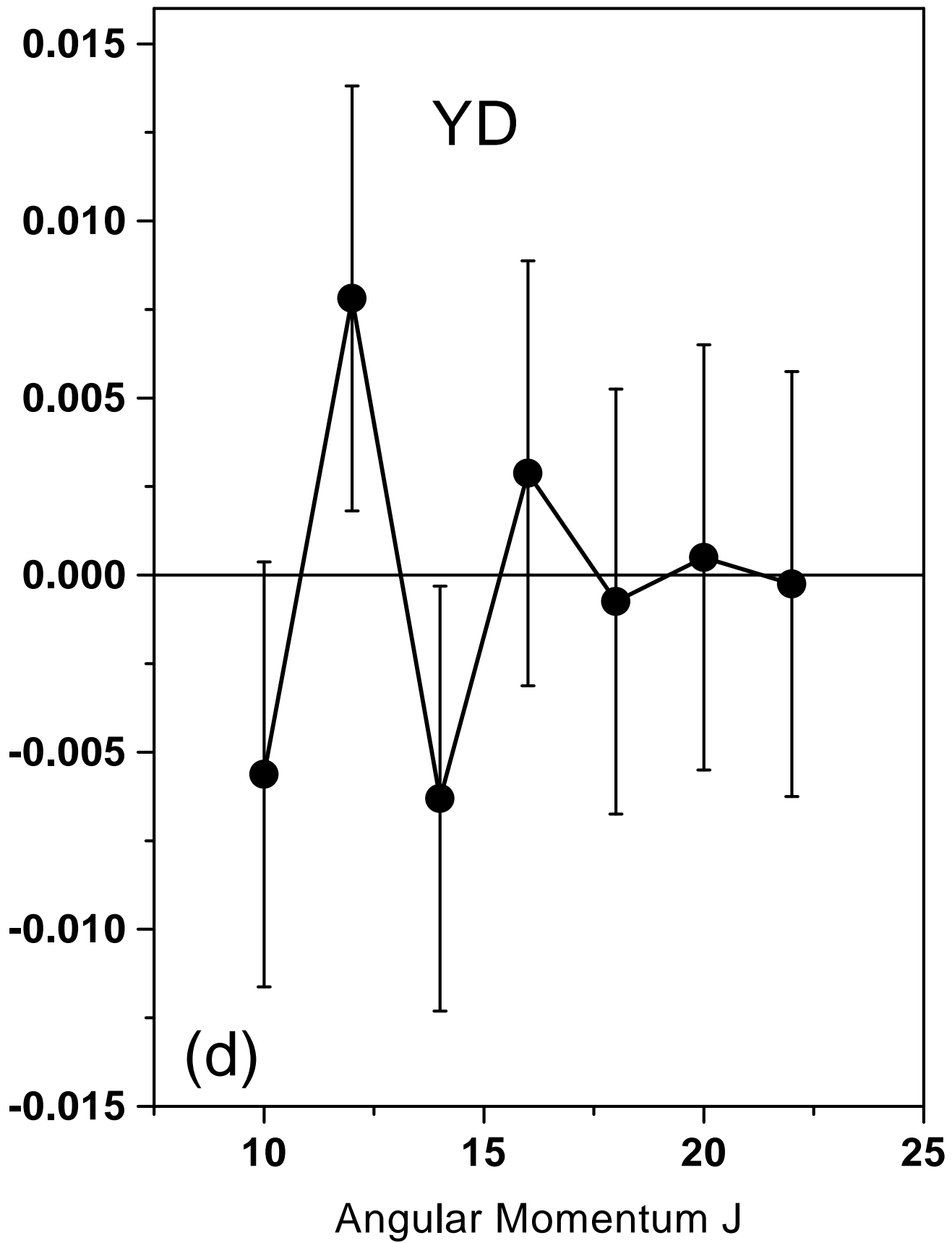
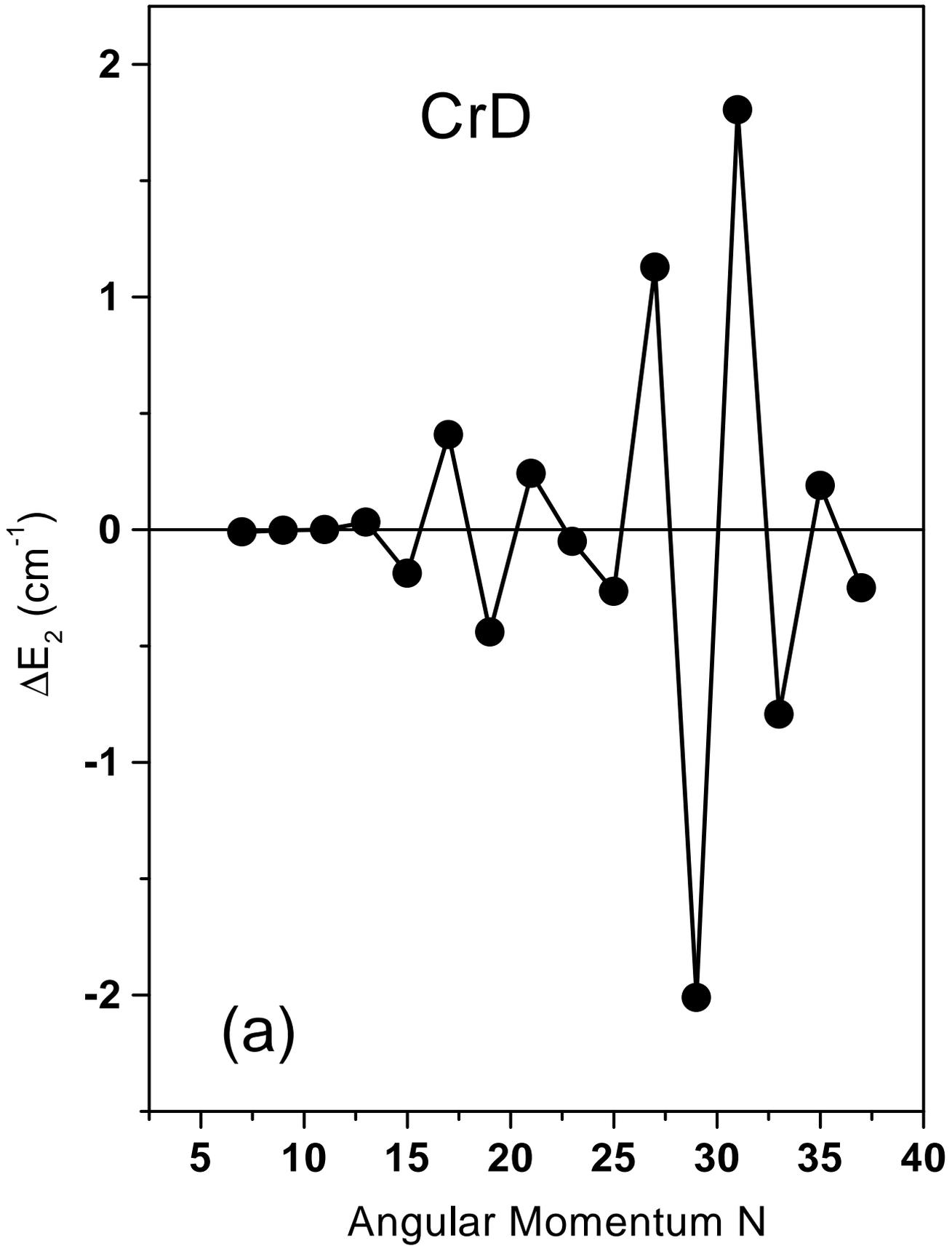
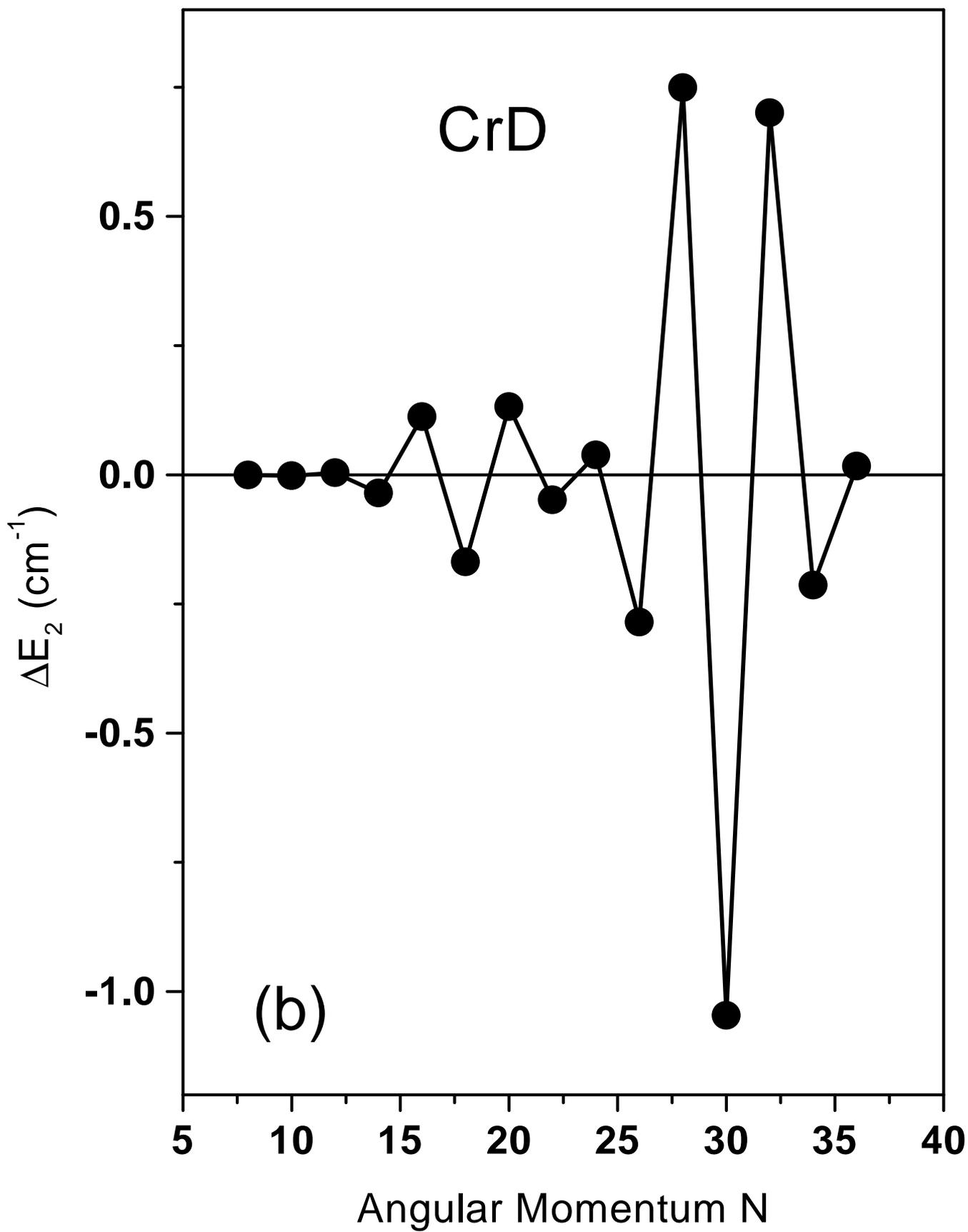
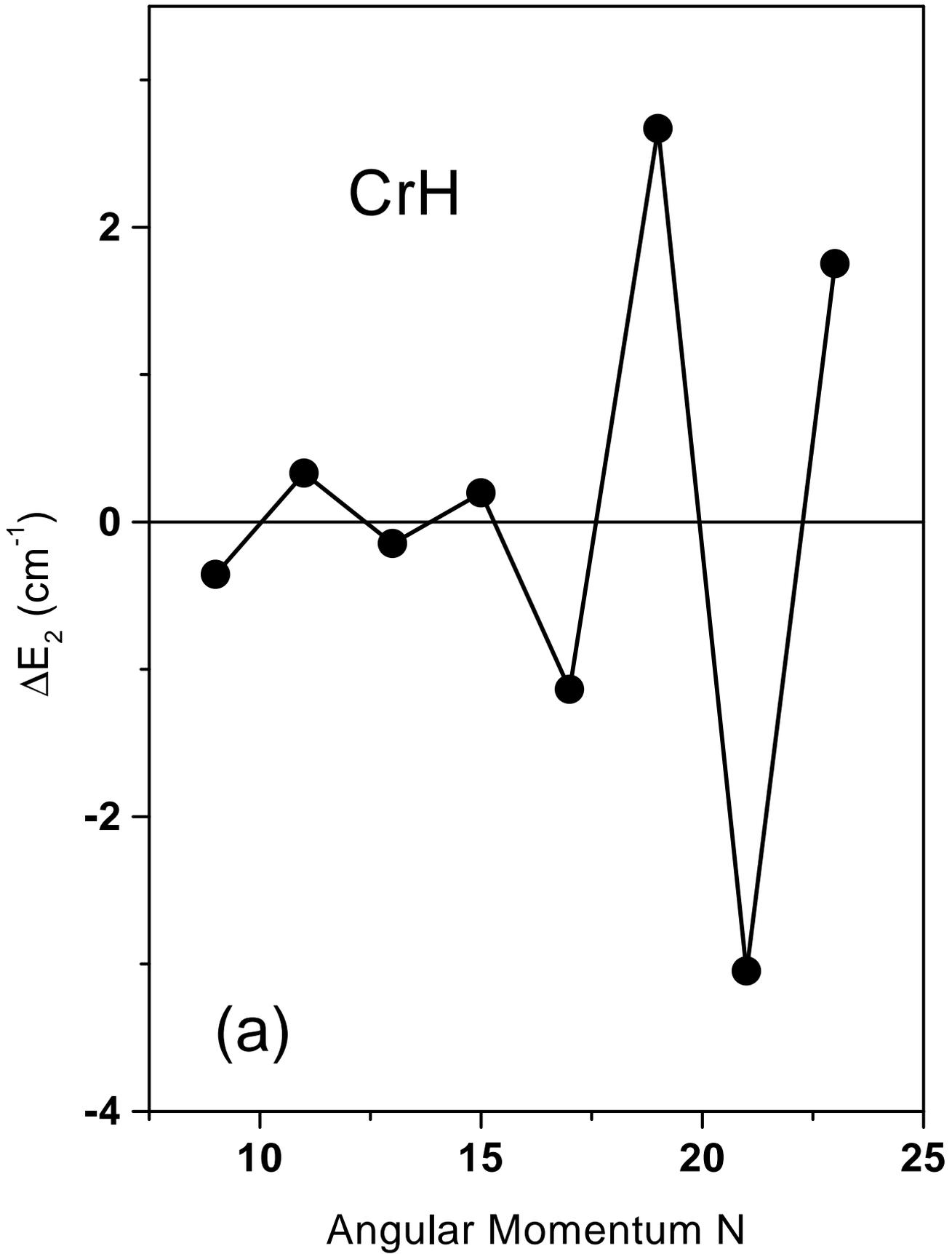
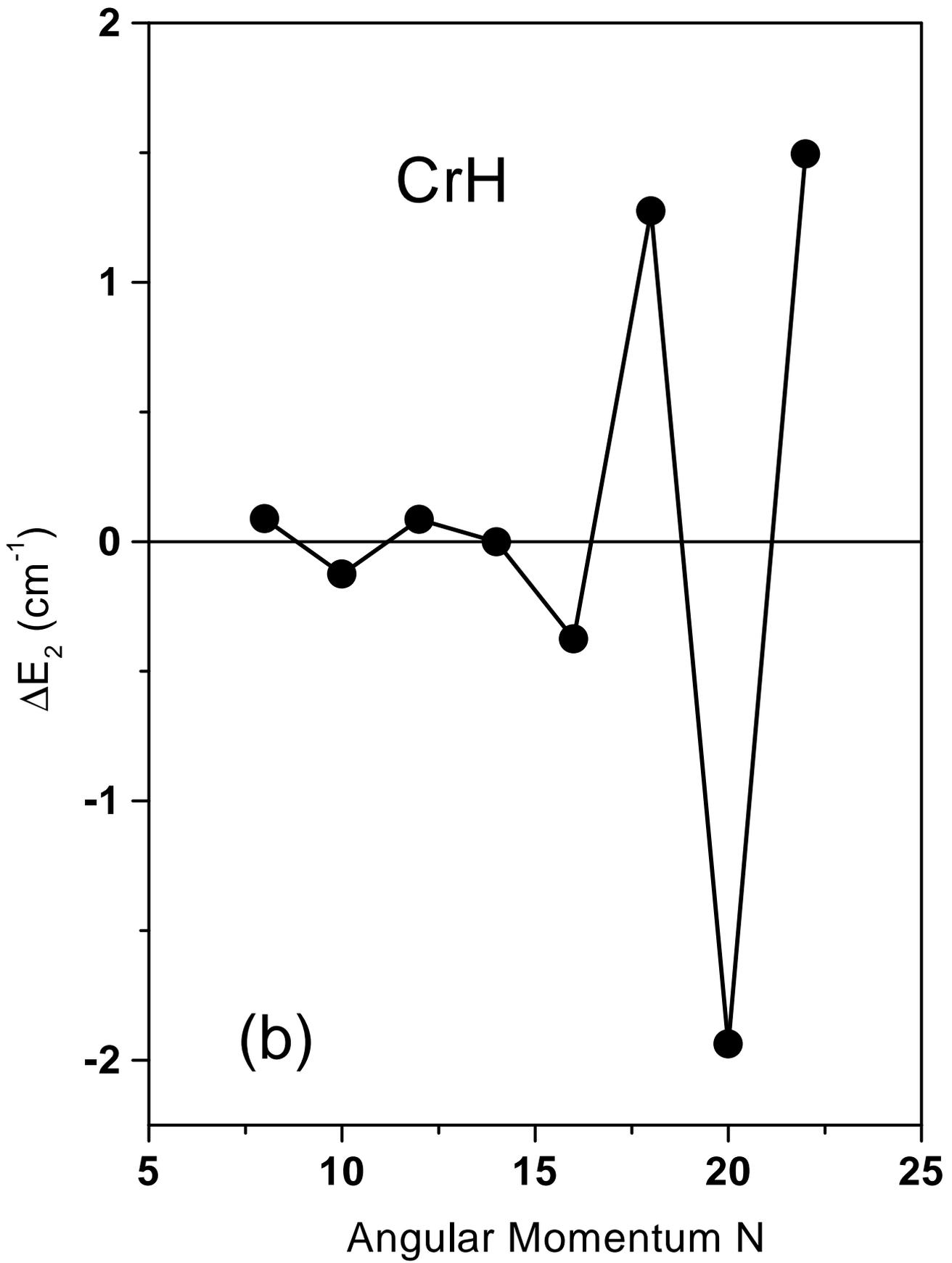
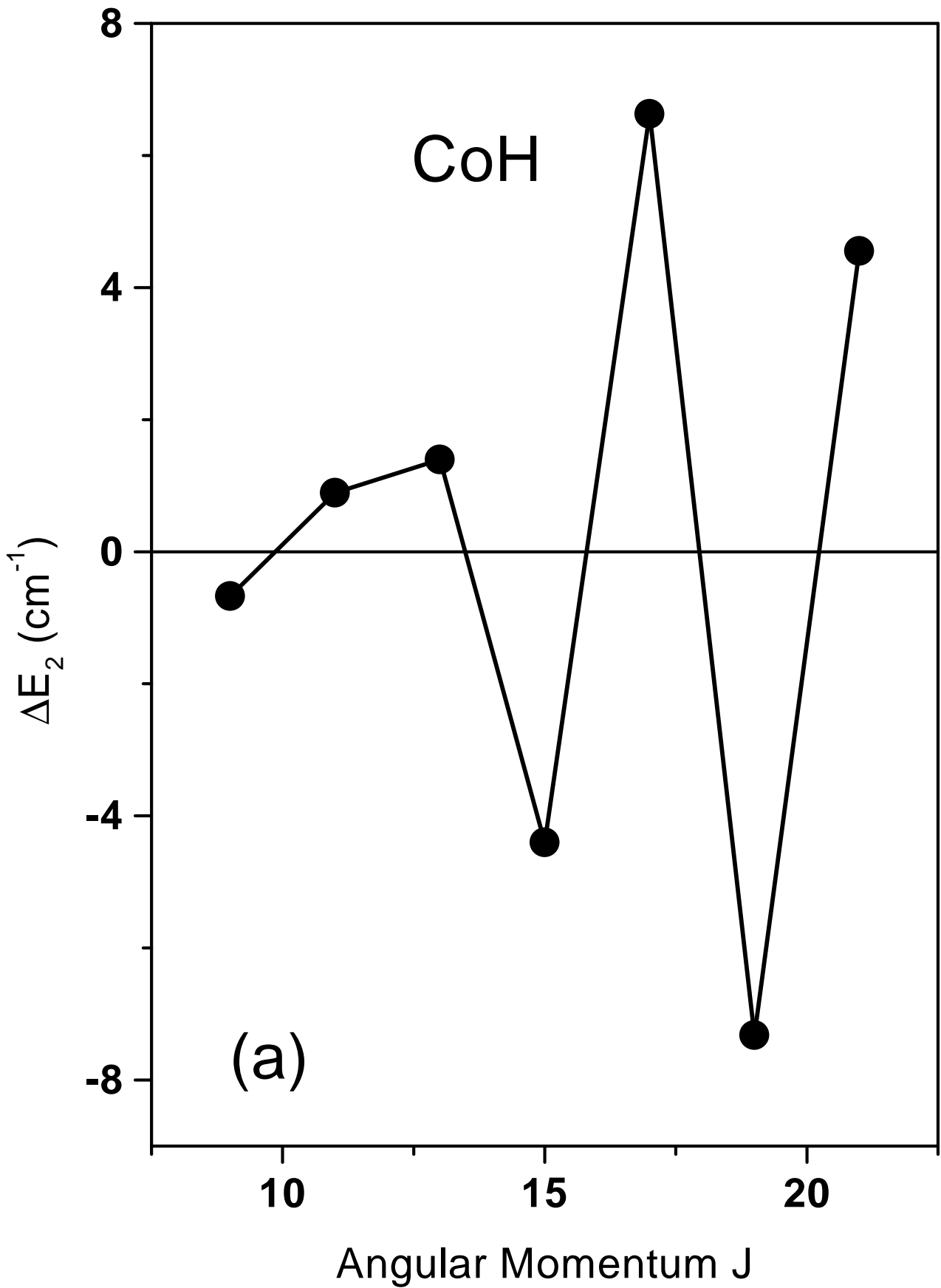
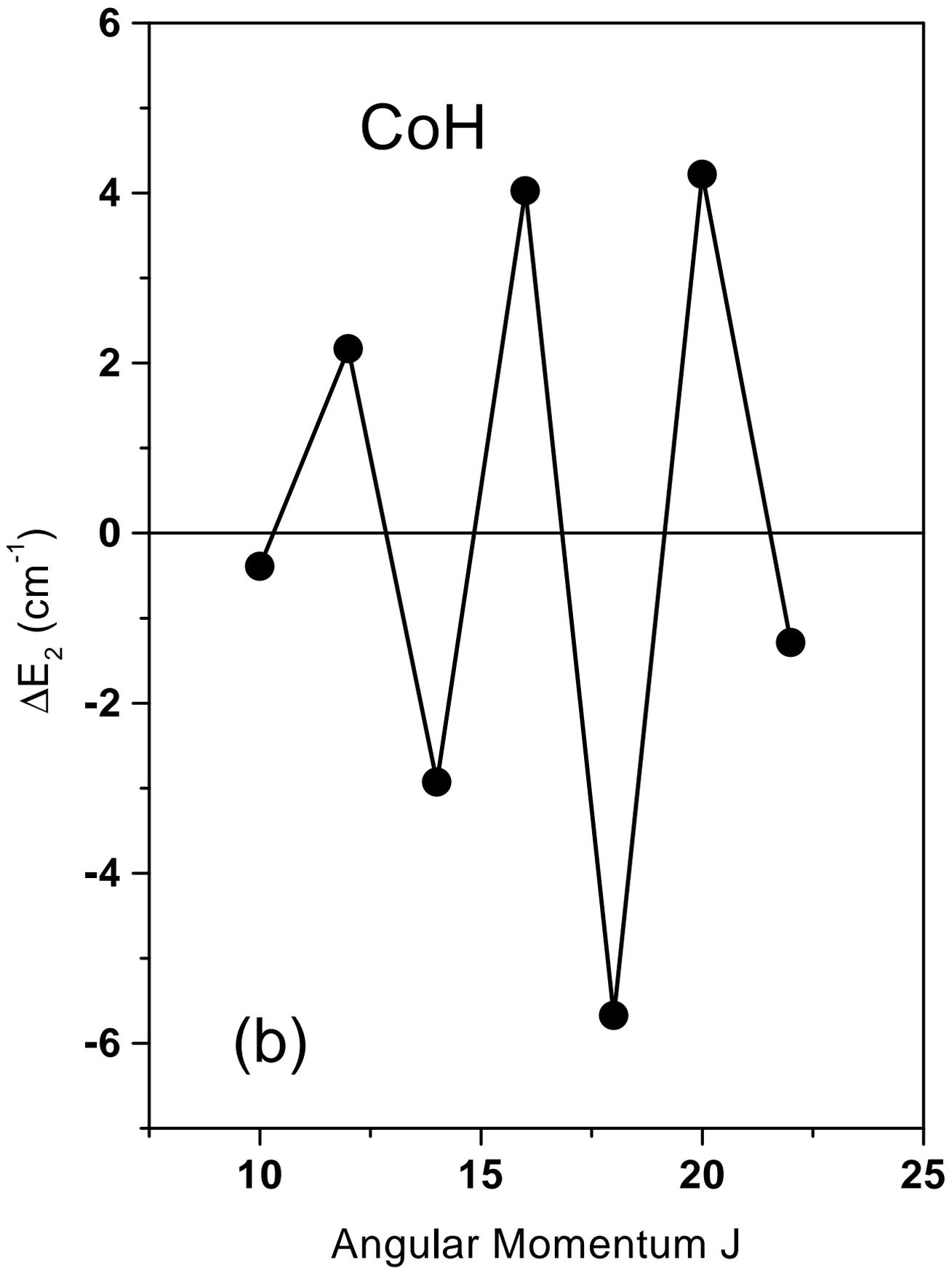
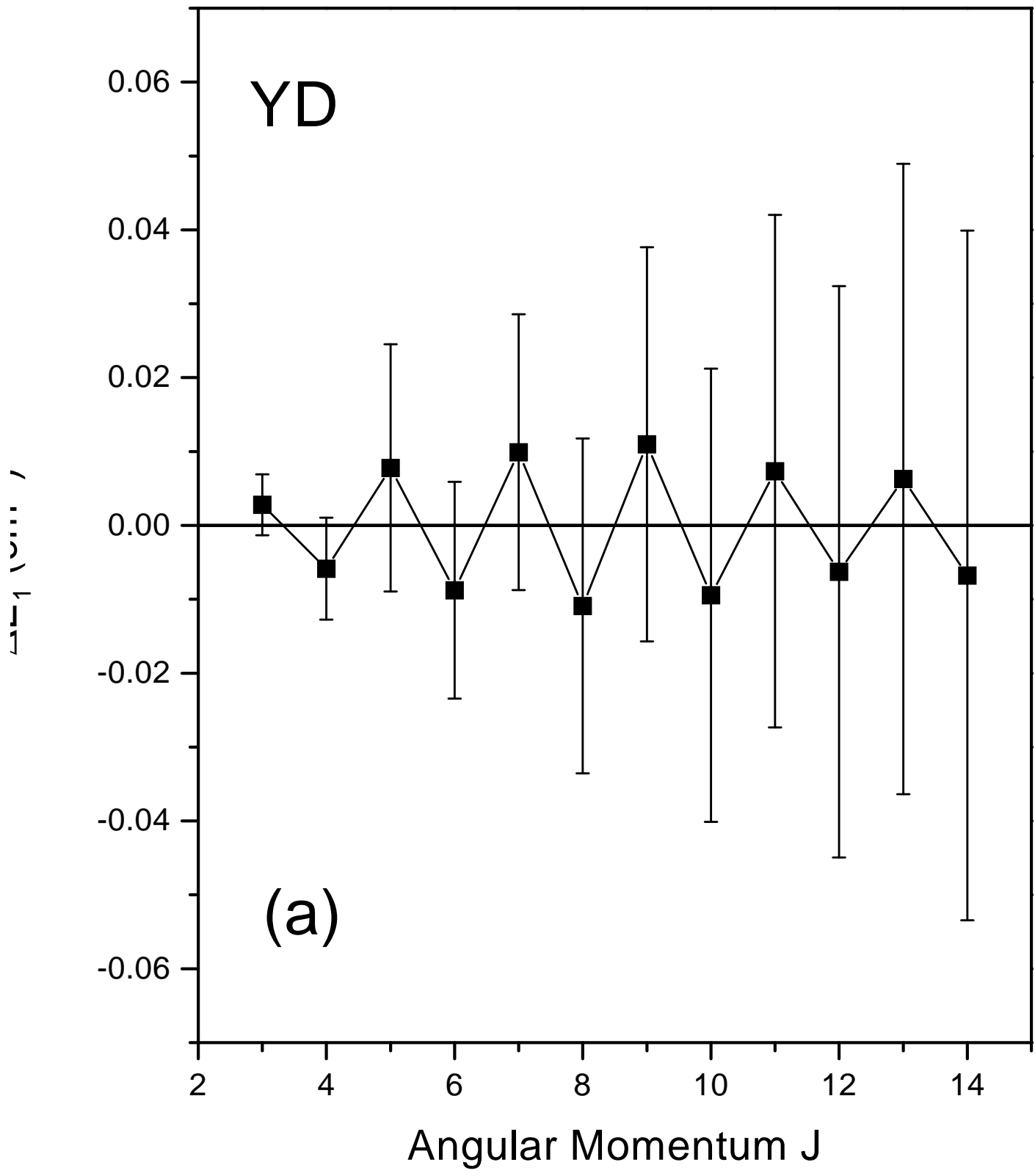
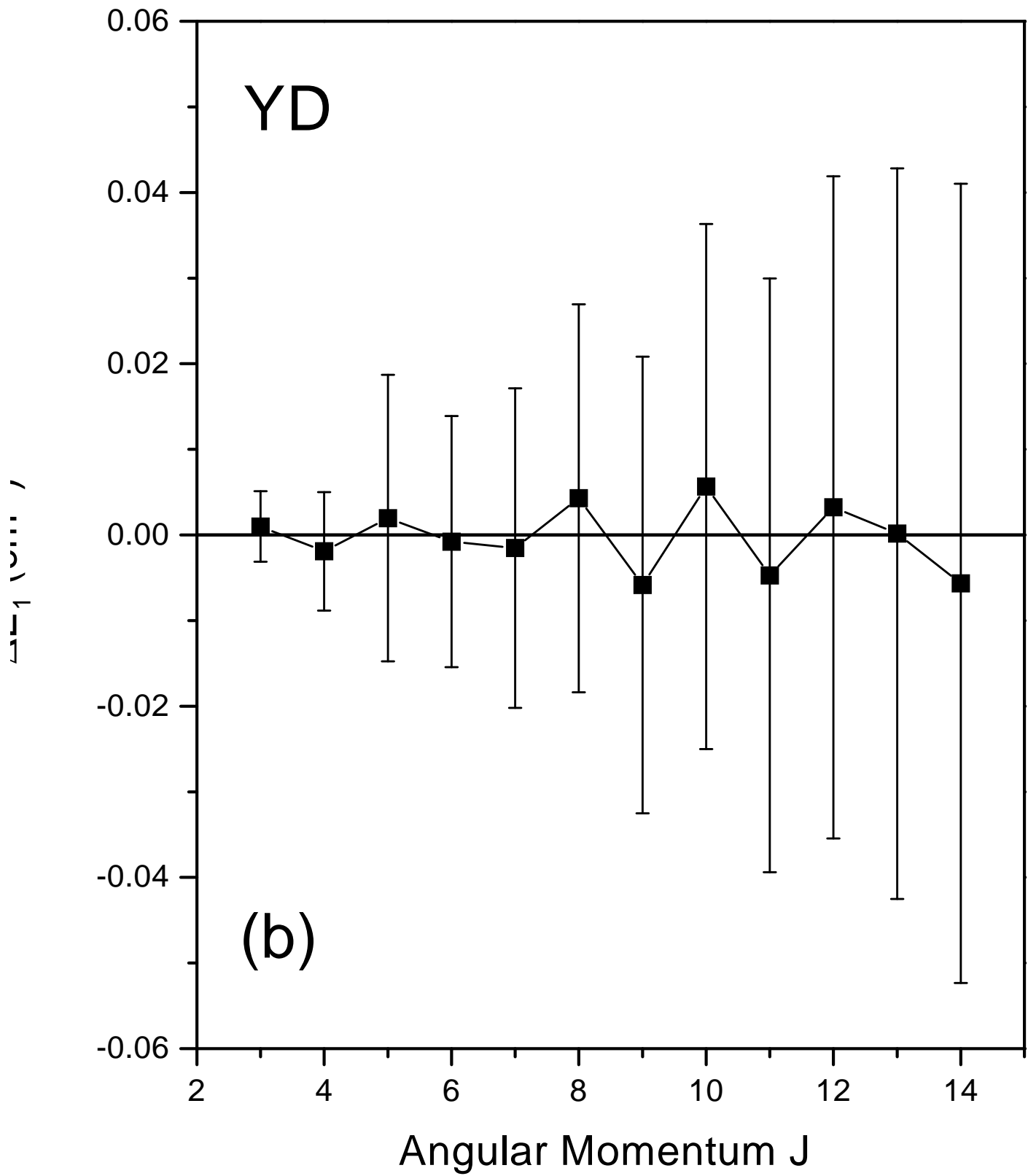
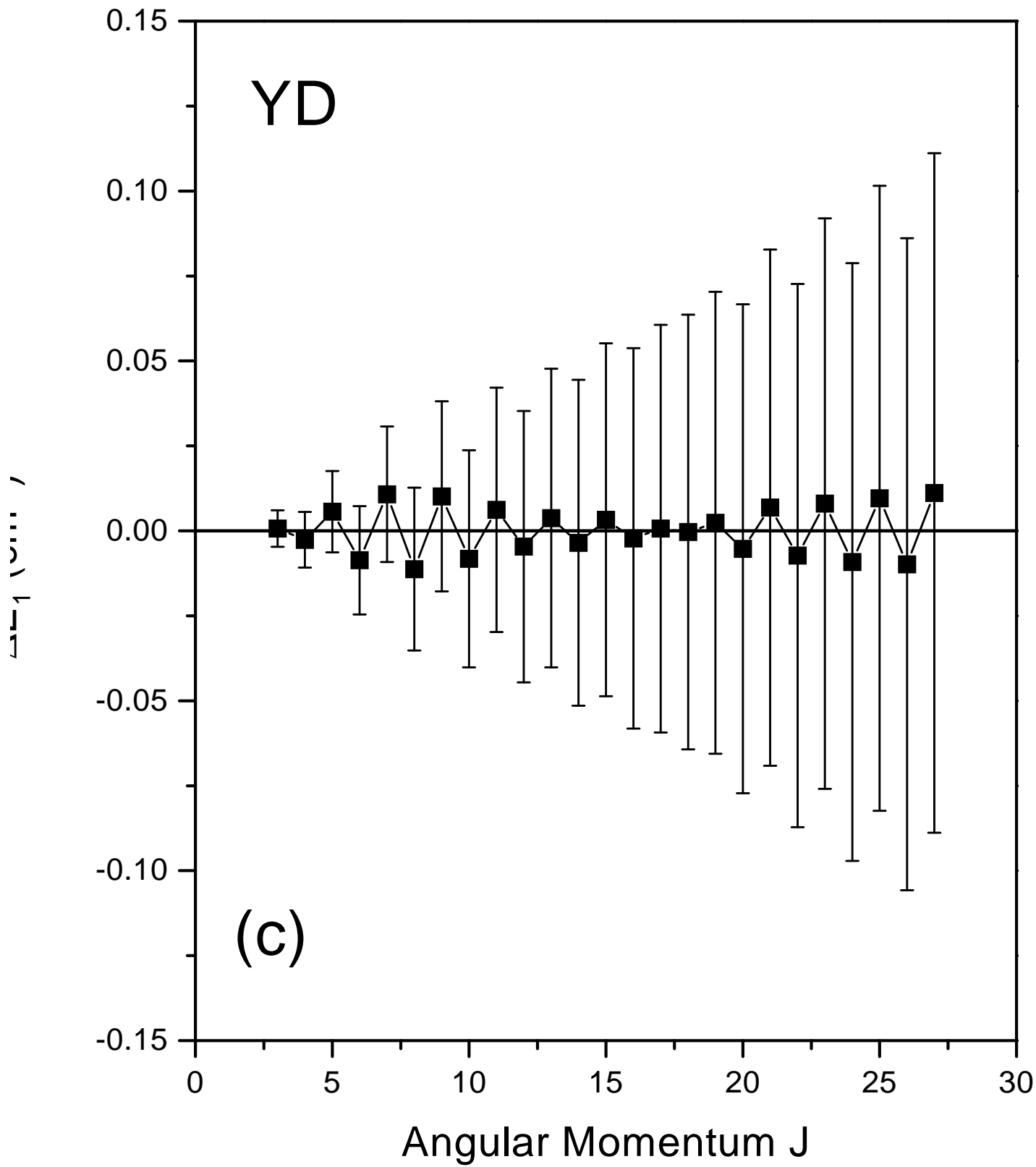
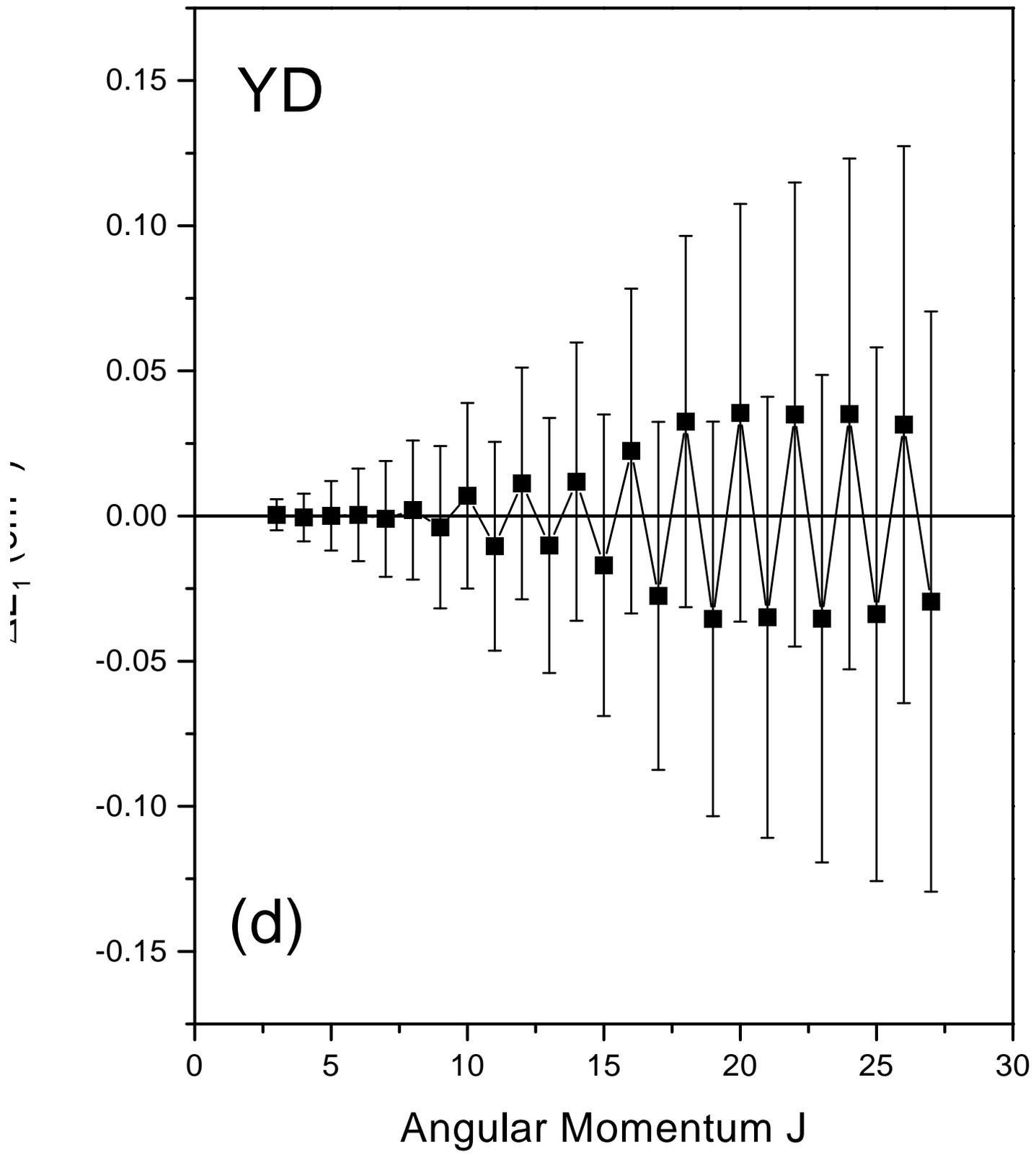
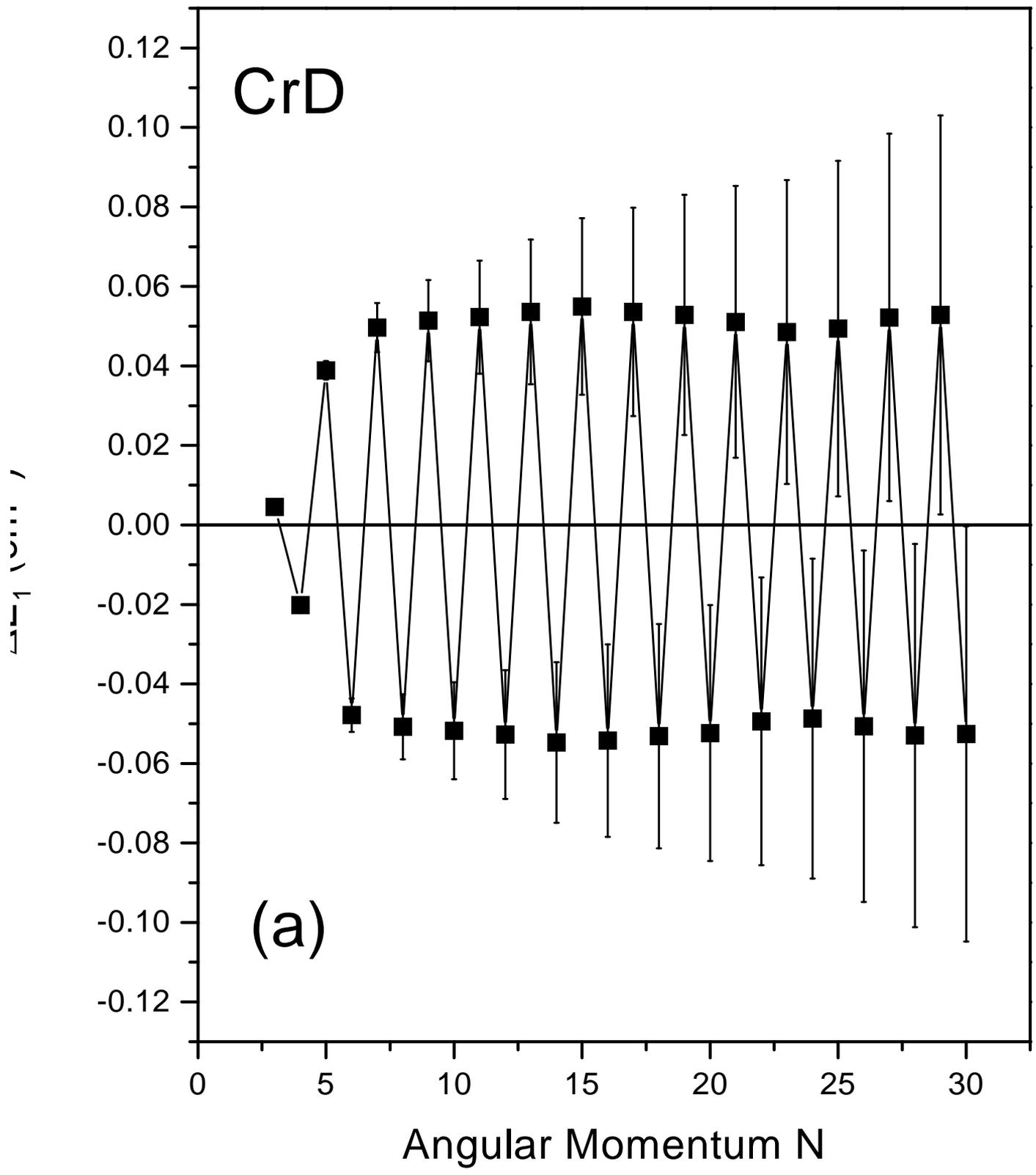
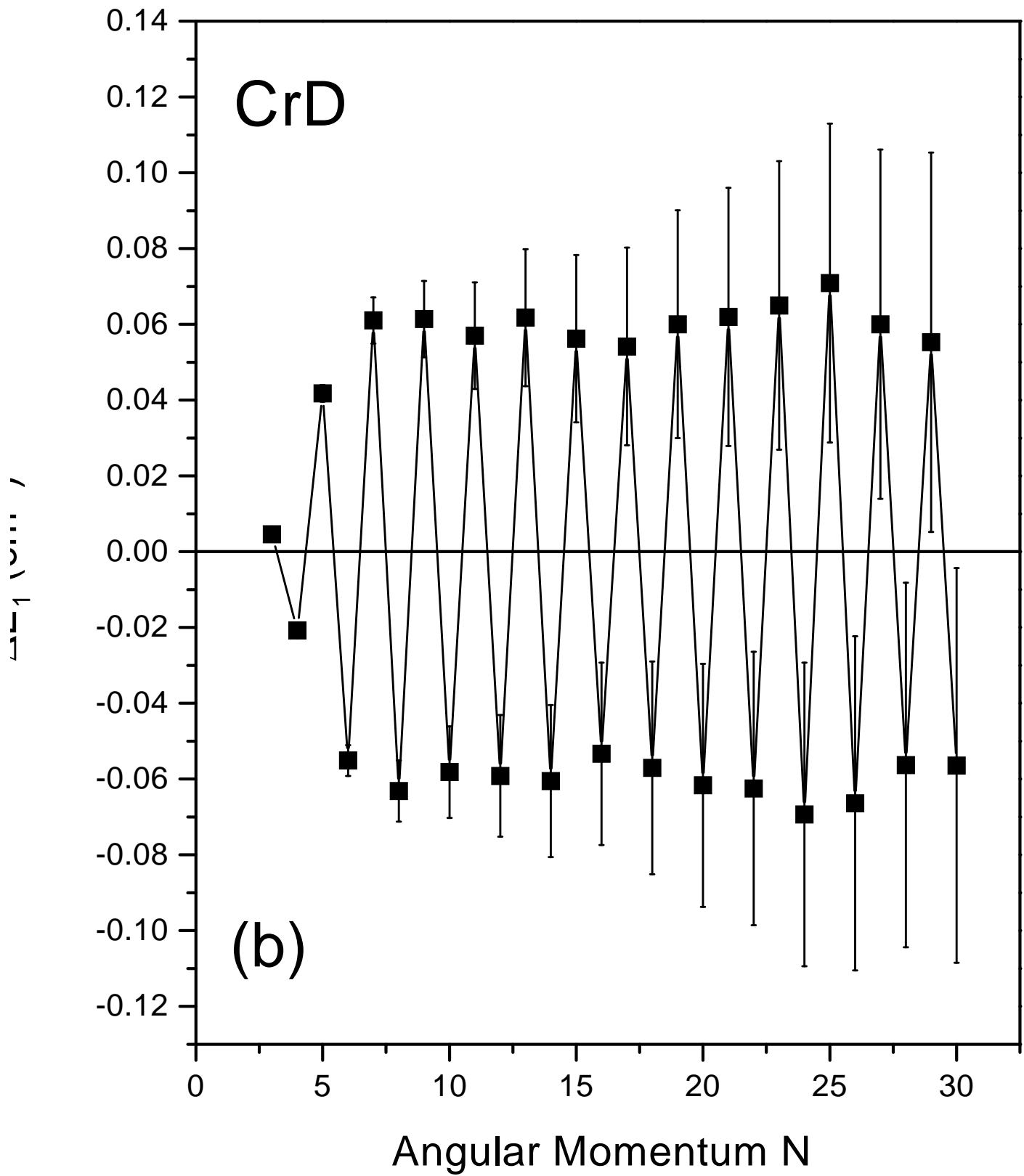
\end{document}